\documentclass[a4paper,12pt]{article}
\usepackage[T2A]{fontenc}
\usepackage[utf8]{inputenc}   
\usepackage[russian,english]{babel}

\usepackage{verbatim} 	
\usepackage{amsmath, amsthm}
\usepackage{graphicx}
\usepackage{array} 	

\usepackage{rotating}	

\usepackage{hyperref}	

\renewcommand{\thefootnote}{\fnsymbol{footnote}}

\begin{document}
\pagestyle{plain}
\pagenumbering{arabic}

\selectlanguage{english}

\title{Doping effects in AlGaAs lasers with separate confinement heterostructures (SCH). 
Modeling optical and electrical characteristics with Sentaurus TCAD.}

\date{}

\author{Z. Koziol\footnote{Corresponding author email: zbigniew@ostu.ru}, and S. I. Matyukhin,\\
Orel State Technical University,\\
29 Naugorskoye Shosse, Orel, 302020, Russia.}

\maketitle

\renewcommand{\thefootnote}{\arabic{footnote}}

\begin{abstract}

Optical and electrical characteristics of AlGaAs lasers with separate confinement heterostructures
are modeled by using Synopsys's Sentaurus TCAD and open source software for 
semi-automatic data analysis of large collections of data. The effects of doping in all laser layers 
are investigated with the aim to achieve optimal characteristics of the
devise. The results are compared with these obtained for real lasers produced at Polyus 
research institute in Moscow, showing that a significant improvement can be achieved, in particular 
an increase in optical efficiency (up to over 70 \%) by careful control of type and level of doping through 
out the entire structure.

\end{abstract}

\clearpage

\baselineskip=3.00ex 


\tableofcontents


\section{Introduction}

An idea of Alferov et al., \cite{Alferov}, 
comprising the use of a geometrically-narrow active recombination region 
where photon generation occurs (Quantum Wells; QW),
with waveguides around improving the gain to loss ratio
(separate confinement heterostructures; SCH), has largely dominated the field of optoelectronics development
in the past years. AlGaAs edge emitting lasers are an example of practical realization of these ideas.

Now, they are mostly used for pumping solid state $Nd:YAG$ lasers used either in high-power metallurgical 
processes or, already, in first field experiments as a highly directional source of energy 
in weapons interceptors.

In our earlier works we first were able to find agreement between our calculations of quantum 
well energy states and the lasing wavelength observed experimentally \cite{Koziol}. 
Next \cite{Waveguide}, several changes in structure of SCH AlGaAs lasers 
have been shown to considerably improve their electrical and optical parameters. 
We compared computed properties with these of lasers produced by Polyus research 
institute in Moscow \cite{Andrejev}, \cite{Andrejev_2}.

In particular, by changing the width of active region (Quantum Well), waveguide width, 
as well by changing the waveguide profile by introducing a gradual
change of Al concentration, we were able to decrease significantly the lasing threshold current, 
increase the slope of optical power versus current, and increase optical efficiency \cite{Waveguide}. 

We have shown also \cite{NGC} that the lasing action may not occur at certain widths
or depths of Quantum Well (QW), and the threshold current as a function of the width may have 
discontinuities. The effects are more pronounced at low temperatures. We argue that these 
discontinuities occur when the most upper quantum well energy values are very close to either 
conduction band or valence band energy offsets. The effects may be observed at certain conditions 
in temperature dependence of lasing threshold current as well.

The main chalanges with the broader use of AlGaAs based SCH lasers are related to improving
some of their electro-optical characteristics, in particular their optical efficiency.

The purpose of this work is to investigate the role of doping levels across the laser structure, and,
if only possible, to find doping concentrations that would lead to the best opto-electrical
parameters, maximizing optical efficiency. 

We assume here uniform doping profiles accross laser layers.

For simulations, we use Sentaurus TCAD from Synopsys \cite{tcad}, which is an advanced commercial 
computational environment, a collection of tools for performing modeling of electronic devices. 

\section{Lasers structure and callibration of modeling.}

We model a laser with $1000  \mu m$ cavity length and $100  \mu m$
laser width, with doping/Al-content as described in Table \ref{table_1}.
The table \ref{table_2} describes its experimental parameters.

Synopsys's Sentaurus TCAD, used for modeling,
can be run on Windows and Linux OS. Linux, once mastered, offers more ways of an efficient solving 
of problems by providing a large set of open source tools and ergonomic environment for their use, making
it our preferred operating system.
This is an advanced, flexible set of tools used for modeling a broad range of technological and physical 
processes in the world of microelectronics phenomena. 
In case of lasers, some calculations in Sentaurus have purely phenomenological nature. 
The electrical and optical characteristics depend, primarily, on the following
computational parameters that are available for adjusting:

$AreaFactor$ of electrodes, $A_e$,  $AreaFactor$ in Physics section, $A_{ph}$, 
electrical contact resistance $R_x$. 
There are several parameters for adjustment that are related to microscopic
physical properties of materials or structures studied. However, often their values are either 
unknown exactly or finding them would require quantum-mechanical modeling of electronic band structure
and transport, based on first-principles. This is however not the aim of our work.

In order to find agreement between the calculated results and these observed experimentaly 
(the threshold current $I_{th}$ and the slope of Optical Power, $S=dL/dI$, are such most basic 
laser parameters), we adjust accordingly values of $A_e$ and $A_{ph}$. 

The results for $I_{th}$ and $S$, in this paper, are all shown normalized by $I_{th}^0$ and $S_0$, 
respectively, which are the values computed for the reference laser described in Table \ref{table_1}.

We neglect here the effect of contact resistance, $R_x$, 
by not including buffer and substrate layers and contacts into calculations
(compare with structure described in Table \ref{table_1}). 
We use $InnerVoltage$ parameter available in Sentaurus 
and treat it as a physical quantity that is closely related to voltage applied. 
Another parameter available in Sentaurus, $OuterVoltage$, is simply related to $InnerVoltage$ by 
the ohmic formula: $OuterVoltage = InnerVoltage + R_x \cdot I$. Hence, any results shown
here may be easy adjusted after calculations by adding the effect of $R_x$.

Let us estimate the value of $R_x$. We assume that this is the sum of electrical resistance
of n-substrate, n-buffer, and p-contact layers (see Table \ref{table_1}). By using simplified
formula for each of these layers, $\rho = 1/(n e \mu)$, and electron/hole mobility from database
of Synopsys, we have the following specific resistivity values for these layers (indexed as rows in Table \ref{table_1}):
$\rho_1 = \frac{1}{2720} \frac{cm V}{A}$, $\rho_2 = \frac{1}{1360} \frac{cm V}{A}$, and
$\rho_8 = \frac{1}{2560} \frac{cm V}{A}$. Taking into account appropriate geometrical dimensions,
we obtain: $r_1=0.013 \Omega$, $r_2=3 \cdot 10^{-5} \Omega$, and $r_8=3 \cdot 10^{-5} \Omega$. Hence, 
$R_x$ is dominated by the resistance of n-substrate layer and it is of the order of $R_x = 13 m\Omega$.
At threshold current of $0.1 A$, that small resistance will cause a difference between computed by us 
lasing offset voltage $U_0$ and that one measured by about $1 mV$ only. 
In will however have a noticeable contribution to differential
resistance $dU/dI$.

\clearpage

\begin{table}[t]
\caption{Structure of AlGaAs SCH laser layers used in computer modeling.}
\label{table_1}
\begin{center}
\begin{tabular}{|c|c|c|c|c|c|}
\hline
\bf{No}  &       \bf{Layer}&    \bf{Composition}&    \bf{Doping [$cm^{-3}$]}&    \bf{Thickness [$\mu m$]}\\
\hline
1        & n-substrate     &   n-GaAs (100)    &  $2\cdot 10^{18}$  &    350              \\
\hline
2        & n-buffer        &   n-GaAs           &  $1\cdot 10^{18}$ &    0.4              \\
\hline
3        & n-emitter       &   $Al_{0.5}Ga_{0.5}As$ & $1\cdot 10^{18}$ &    1.6              \\
\hline
4        & waveguide       &   $Al_{0.33}Ga_{0.67}As$ & none ($n \approx 10^{15}$) &  0.2                \\
\hline
5        & active region (QW) & $Al_{0.08}Ga_{0.92}As$ & none ($n \approx 10^{15}$) &  0.012                \\
\hline
6        & waveguide       &   $Al_{0.33}Ga_{0.67}As$ & none ($n \approx 10^{15}$) &  0.2                \\
\hline
7        & p-emitter       &   $Al_{0.5}Ga_{0.5}As$ &  $1\cdot 10^{18}$ &  1.6        \\
\hline
8        & contact layer   &   p-GaAs           & $4\cdot 10^{19}$ &    0.5      \\
\hline
\end{tabular}
\end{center}
\end{table}

\begin{table}[t]
\caption{Summary of experimental conditions and laser parameters.}
\label{table_2}
\begin{center}
\begin{tabular}{|c|c|c|c|c|c|}
\hline
Temperature [K] &   300       \\
\hline
Lasing wavelength [nm]&   808 \\
\hline
Offset voltage $U_0$ [V] &   1.56-1.60\\
\hline
Differential resistance, $r=dU/dI$  [m$\Omega$] &   50-80   \\
\hline
Threshold current $I_{th}$ [mA] & 200-300 \\
\hline
Slope of optical power, $S=dL/dI$ [W/A]&  1.15-1.25 \\
\hline
Left mirror reflection coefficient $R_l$ &  0.05 \\
\hline
Right mirror reflection coefficient $R_r$ &  0.95 \\
\hline
\end{tabular}
\end{center}
\end{table}

\clearpage

\section{Methods of data analysis.}

\subsection{Using Perl, TCL, gnuplot, and other open source tools on Linux.}

It is possible to work in a batch mode in Sentaurus TCAD. 
However, we find it more convenient to use Perl\footnote{Perl stands for \emph{Practical Extraction and 
Report Language}; http://www.perl.org} scripts rather for control 
of batch processing and changing parameters of calculations. It is a very flexible programming language, 
suitable in particular for working on text files (e.g. manipulation on text data files), 
and it is convenient to be used
from terminal window rather, not by using a GUI interface, which 
is a more productive approach towards computation. 

Tcl\footnote{\emph{Tool Command Language}; http://www.tcl.tk} is a very powerful but easy to 
learn dynamic programming language, suitable for a wide range of uses.
Sentaurus TCAD contains libraries designed to be used together with TCL. 
They are run through Sentaurus's tdx interface and are used for manipulating (extracting) 
spacial data from binary TDR files. 

Besides, we use tools/programs like gnuplot, grep, shell commands, etc. A more detailed 
description, with examples of scripts, is available 
on our laboratory web site\footnote{http://www.ostu.ru/units/ltd/zbigniew/synopsys.php}.

\subsection{Threshold current and $L(I)$ dependence.}

The most accurate way of finding $I_{th}$ is by extrapolating the linear part of $L(I)$ 
to $L=0$ just after the current larger than $I_{th}$. We used a set of gnuplot and perl scripts for that
that could be run semi-automatically on a large collection of data.
One should only take care here that a properly chosen is the data range for fiting,
since $L(I)$ is a linear function in a certain range of $I$ values only. The choice of that
range may affect accuracy of data analysis. We find however that this is the most accurate 
effective method to analyse the data from a large collection of datasets.

\subsection{Three ways of finding lasing offset voltage $U_0$.}

\subsubsection{$U_0$ from fiting $U(I)$ dependence}

Textbooks' exponential $U(I)$ dependence is fullfiled well at voltages which are well below
the lasing offset voltage $U_0$. Near the lasing threshold, we observe 
a strong departure from that dependence, 
and, in particular, for many data curves a clear kink in $U(I)$ is observed at $U_0$. 
We find that a modified exponential dependence describes the data very well:
\begin{equation}\label{exponential_6_parameters}
\begin{array}{ll}
	I(U) = I_{th} \cdot exp(B\cdot (U-U_0) + C \cdot (U-U_0)^2),~~~ for ~ U< U_0\\
	I(U) = I_{th} \cdot exp(D\cdot (U-U_0) + E \cdot (U-U_0)^2),~~~ for ~ U> U_0
\end{array}
\end{equation}

where $I_{th}$, $U_0$, as well $B$, $C$, $D$, and $E$ are certain fiting parameters.

\begin{figure}[t]
\begin{center}
      \resizebox{150mm}{!}{\includegraphics{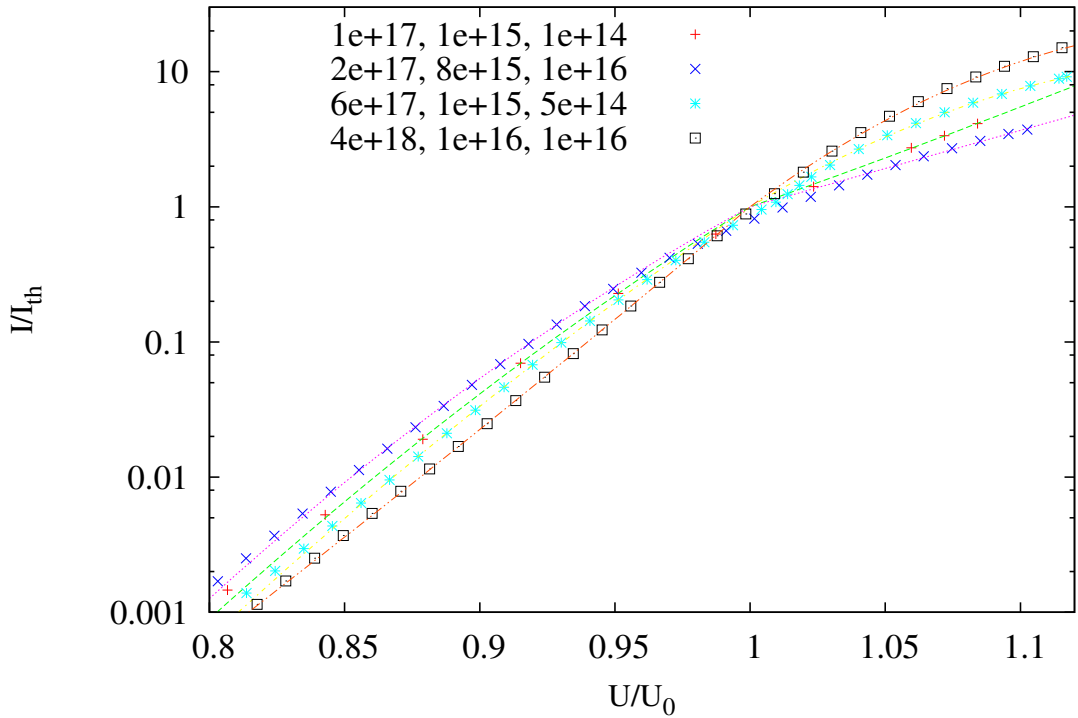}}
      \caption{Examples of typical $I-V$ characteristics for a few combination (as described
in the Figure) of doping concentrations (n- and p-emitters concentration first, followed by waveguides
and active region concentrations). We show the narrow region near the lasing threshold, only.
The curves are computed by using fiting parameters ($I_{th}, B, C, D, E, U_0$) of equation \eqref{exponential_6_parameters},
and after that voltage is normalized by $U_0$ and current by $I_{th}$. 
}
      \label{doping21a}
\end{center}
\end{figure}

The above function is continuous at $U_0$, as it obviously should, 
but it's derivative is usually not. We used Equation (\ref{exponential_6_parameters}) to find out $I_{th}$
and $U_0$. However, the accuracy of this method was found lower than accuracy of the following 
two methods described. Figure \ref{doping21a} shows a few typical examples of I(U) dependencies,
together with lines computed to fit them by using (\ref{exponential_6_parameters}).

Let us notice that since (\ref{exponential_6_parameters}) may have a discontinuous derivative,
using it to find out differential resistance at $U_0$ is ambiguous. 
From Equation (\ref{exponential_6_parameters}), at $U=U_0$, we will have $dU/dI=\frac{1}{I_{th} \cdot B}$
on the side $U<U_0$ and $dU/dI=\frac{1}{I_{th} \cdot D}$ on the side $U>U_0$.

In fact, the differential
resistance measured in experiments is, to some extend, determined by the resistance
of buffer layers and contacts, and by the way how it has been actually defined for non-linear $I-V$ curves.

In practice, the $U(I)$ dependence, above $U_0$ but not far above, may be asummed to be a linear function.
In such a case, we find from data anaylysis, for the third dataset, for instance, in Figure \ref{doping21a},
$dU/dI \approx 50 m\Omega$, which, together with computed contact resistance $R_x=13 m\Omega$ gives good
qualitative agreement with the differential resistance expected for real laser, as listed in Table \ref{table_2}.


\subsubsection{$U_0$ from maximum of $dlogL(U)/dU$.}

Another approach to find $U_0$ is by finding position of maximum in derivative of logarithm of $L$ versus voltage:
$d L(U)/(LdU)$. This is a very accurate method when there is a sufficiently large number of datapoints
available near $U_0$. However, for that, we would have to perform a lot of computations that are time
consuming, in small enough steps in $U$. Hence, this method is not always effective.

\subsubsection{$U_0$ from gain versus voltage curves.}

We find that a good accuracy of determining $U_0$ is from extrapolating linearly gain versus voltage 
curve in a near range of voltage values below $U_0$, to the value of maximal gain, which is constant 
above $U_0$. The results presented in this paper were obtained that way.

\subsection{Differential resistance and optical efficiency.}

Let us use the simplified assumption that $U(I)$ is linear above $U_0$: $U(I)= U_0 + r\cdot (I-I_{th})$.
Together with linear dependence of lasing light power versus current, $L=S\cdot (I-I_{th})$, 
we have the following relation between optical power efficiency, $\eta=L/P$, and current $i=(I-I_{th})/I_{th}$:

\begin{equation}\label{kpd_equation}
\begin{array}{ll}
	\eta = \frac {S}{U_0} \cdot \frac {i}{\left( 1 + \frac {r \cdot I_{th}}{U_0} \cdot i \right)(1+i)}
\end{array}
\end{equation}

Equation \ref{kpd_equation} gives a resonably accurate qualitative description of $\eta(i)$. The parameter
determining position and value of maximum, $\eta_{max}(i_{max})$, is governed by factor 
$\alpha=\frac {r \cdot I_{th}}{U_0}$. In a typical case, with $r=60 m\Omega$, $U_0=1.65 V$, and $I_{th}=0.2 A$,
we have $\alpha \approx 7 \cdot 10^{-3}$.

We sometime prefer to use another form of Equation \ref{kpd_equation}, where $\eta$ as a function of voltage
is used:

\begin{equation}\label{kpd_equation_u}
\begin{array}{ll}
	\eta = \frac {S}{U_0} \cdot \frac {u-1}{ u \cdot \left( u + \frac {r \cdot I_{th}}{U_0} -1 \right)}
\end{array}
\end{equation}

where $u=U/U_0$.

\begin{figure}[t]
\begin{center}
      \resizebox{150mm}{!}{\includegraphics{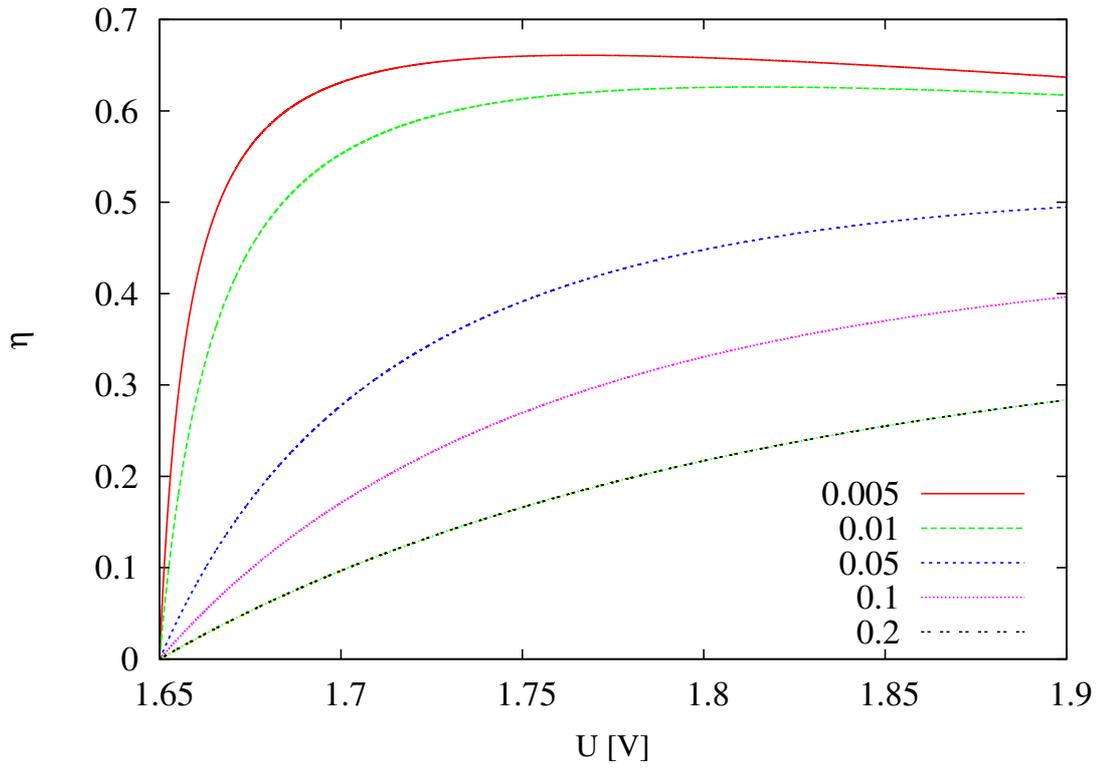}}
      \caption{Optical efficiency as a function of voltage, computed by using Equation \ref{kpd_equation_u},
		for a few values of parameter $\alpha=\frac {r \cdot I_{th}}{U_0}$, as shown in the Figure. 
		It has been assumed that 
		$S=1.25 W/A$ and $U_0=1.65 V$.
}
      \label{eta_01}
\end{center}
\end{figure}

Figure \ref{eta_01} shows a few examples of $\eta(U)$ curves computed with Equation \ref{kpd_equation_u}.


\clearpage

\section{N-N waweguide structure.}

By "N-N" waveguide structure we mean a doping structure as described in Table \ref{table_1},
where both waveguides and the active region (QW) are n-type doped.

We present here selected examples only of data obtained from semi-automatic analyses 
of thousands of datasets. 

\subsection{The role of doping in Active Region (QW).}

\begin{figure}[t]
\begin{center}
      \resizebox{150mm}{!}{\includegraphics{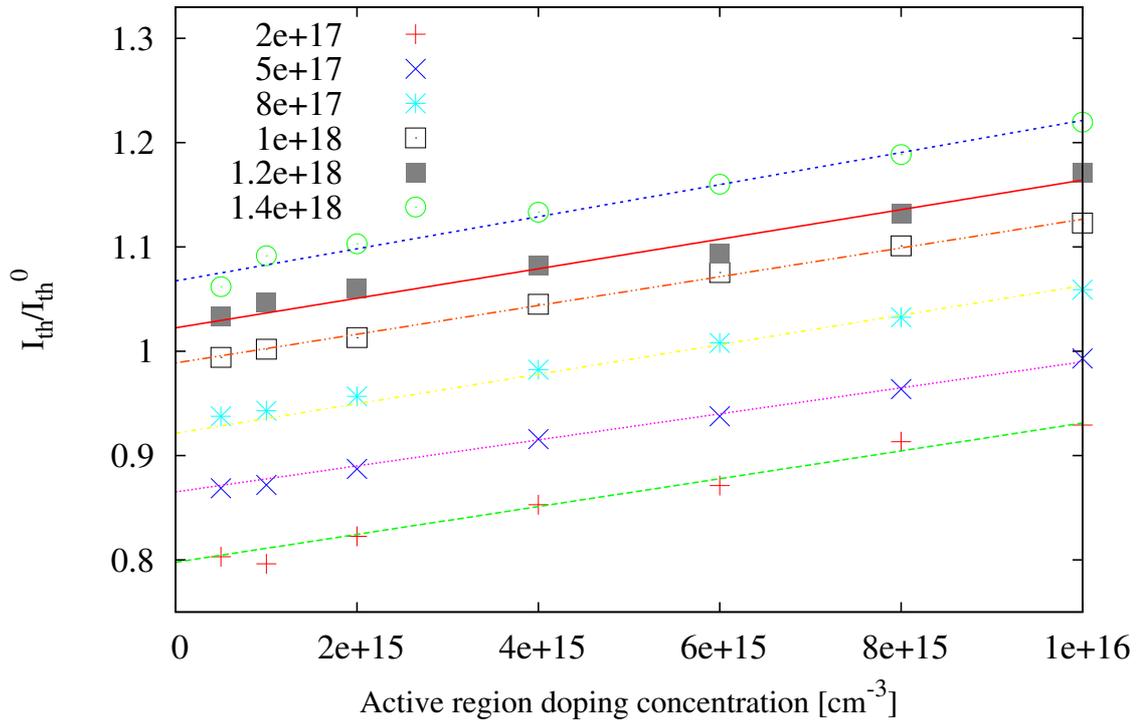}}
      \caption{Threshold current, $I_{th}$, normalized by $I_{th}^0$, as a function of 
doping concentration in active region, for a range of doping concentrations in emitter regions, as
indicated in the figure. Doping concentration in waveguide regions is kept constant 
at value of $10^{15}/cm^3$. 
}
      \label{doping10a}
\end{center}
\end{figure}

\begin{figure}[t]
\begin{center}
      \resizebox{150mm}{!}{\includegraphics{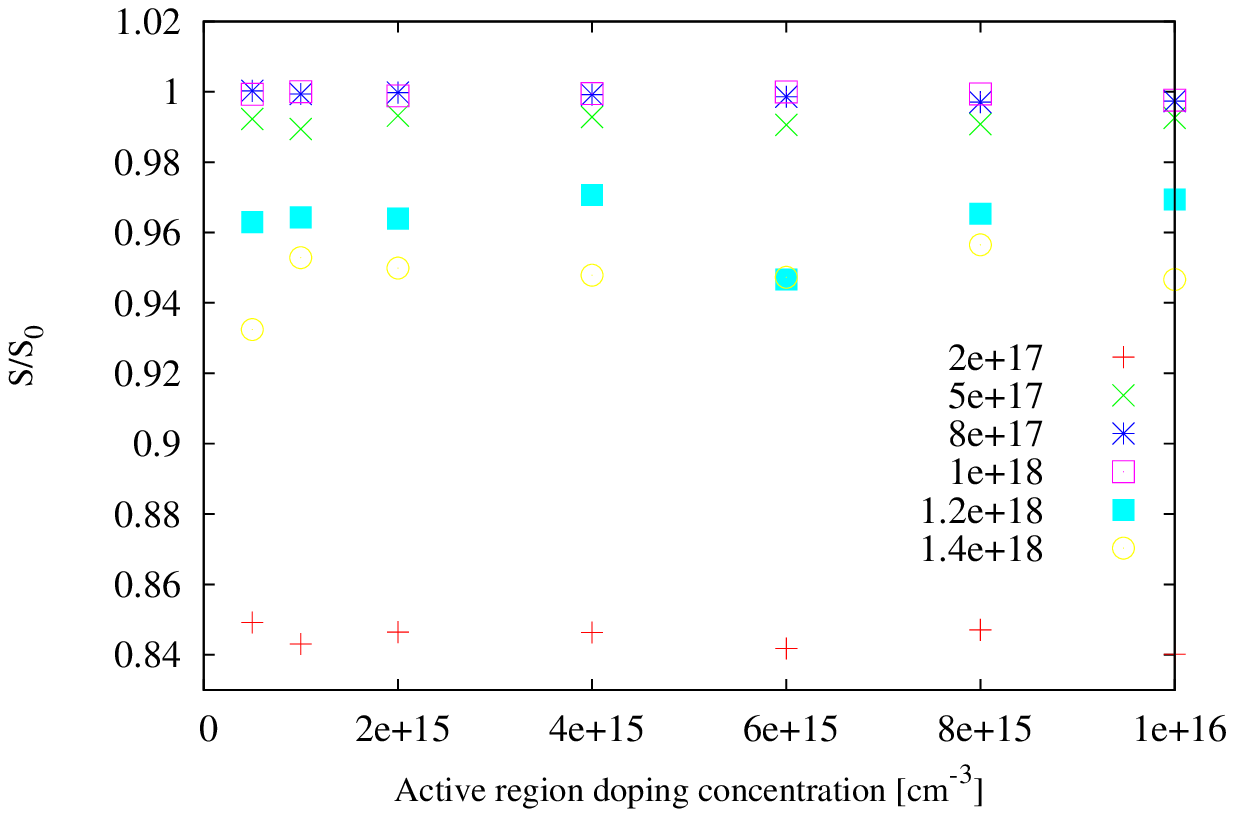}}
      \caption{$S$ normalized by $S_0$, as a function of 
doping concentration in active region, for a range of doping concentrations in emitter regions, as
indicated in the figure. Doping concentration in waveguide regions is kept constant 
at value of $10^{15}/cm^3$. These data correspond to datapoints displayed for $I_{th}$ in Figure \ref{doping10a}.
}
      \label{doping10a_1}
\end{center}
\end{figure}

\begin{figure}[t]
\begin{center}
      \resizebox{150mm}{!}{\includegraphics{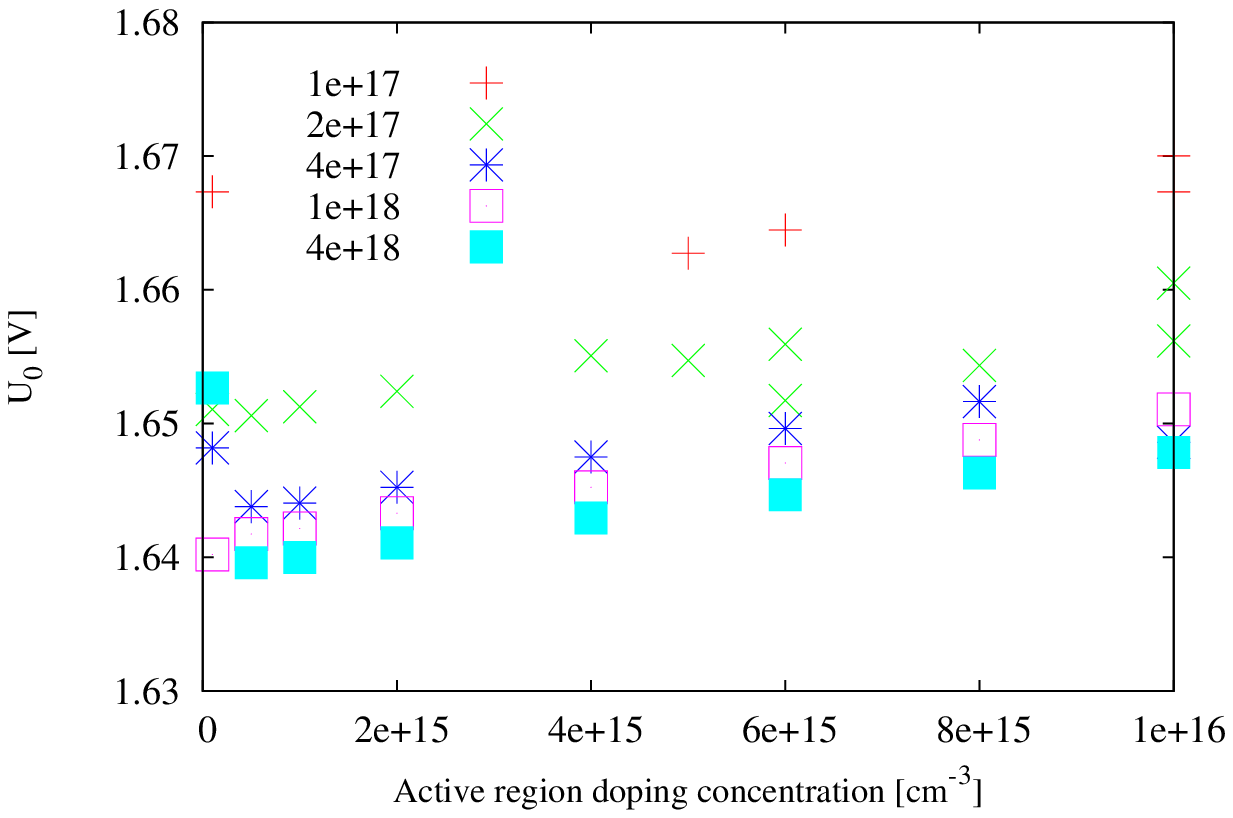}}
      \caption{$U_0$ as a function of 
doping concentration in active region, for a range of doping concentrations in emitter regions, as
indicated in the figure. Doping concentration in waveguide regions is kept constant 
at value of $10^{15}/cm^3$. These data correspond to datapoints displayed for $I_{th}$ in Figure \ref{doping10a}.
}
      \label{doping10a_2}
\end{center}
\end{figure}

Figures \ref{doping10a}, \ref{doping10a_1}, and \ref{doping10a_2} show the role of 
doping concentration in active region, for a few values of doping concentrations 
in emitter regions, on basic characteristics of lasers: threshold current, slope of 
light power versus current, $dL(I)/dI$, and lasing offset voltage, respectively. The 
doping concentration in waveguide regions is the same on these three figures; it is
$10^{15}/cm^3$. 


We see that none of these quantities, i.e. $I_{th}$, $dL/dI$, $U_0$, depends strongly
on active region doping. 

\begin{figure}[t]
\begin{center}
      \resizebox{150mm}{!}{\includegraphics{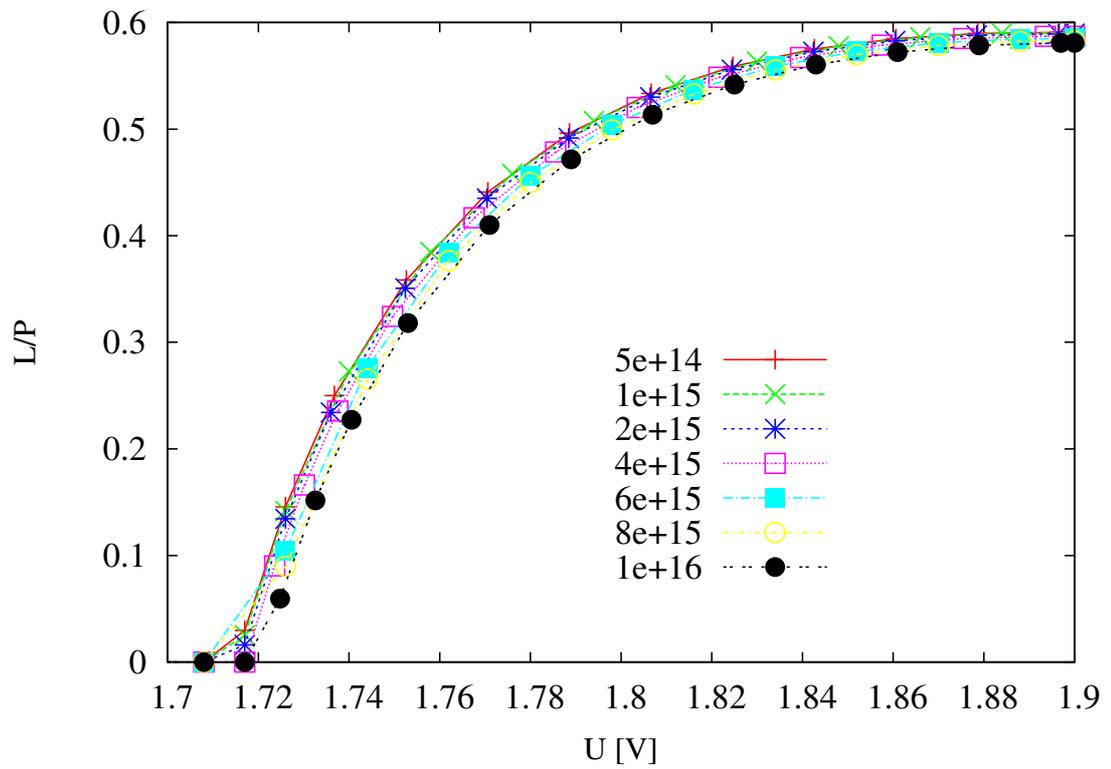}}
      \caption{Optical efficiency versus applied voltage, for a few values
		of doping concentration in active region, as described in the Figure.
		Doping concentrations in waveguide regions is $10^{15}/cm^3$, and in emitters
		regions it is $10^{18}/cm^3$.
}
      \label{kpd01}
\end{center}
\end{figure}

Therefore, it is not unexpected that the optical efficiency, the ratio of optical power $L$ to 
total power supplied to the laser, $P=U\cdot I$, as shown in Figure \ref{kpd01}, weakly 
depends on doping concentration in active region as well. Though for the lower doping the higher
optical efficiency is obtained.

\clearpage

\subsection{The role of doping in Waveguide Regions.}

The effects of doping in waveguide regions are similar to these in the active region, but stronger.

\begin{figure}[t]
\begin{center}
      \resizebox{150mm}{!}{\includegraphics{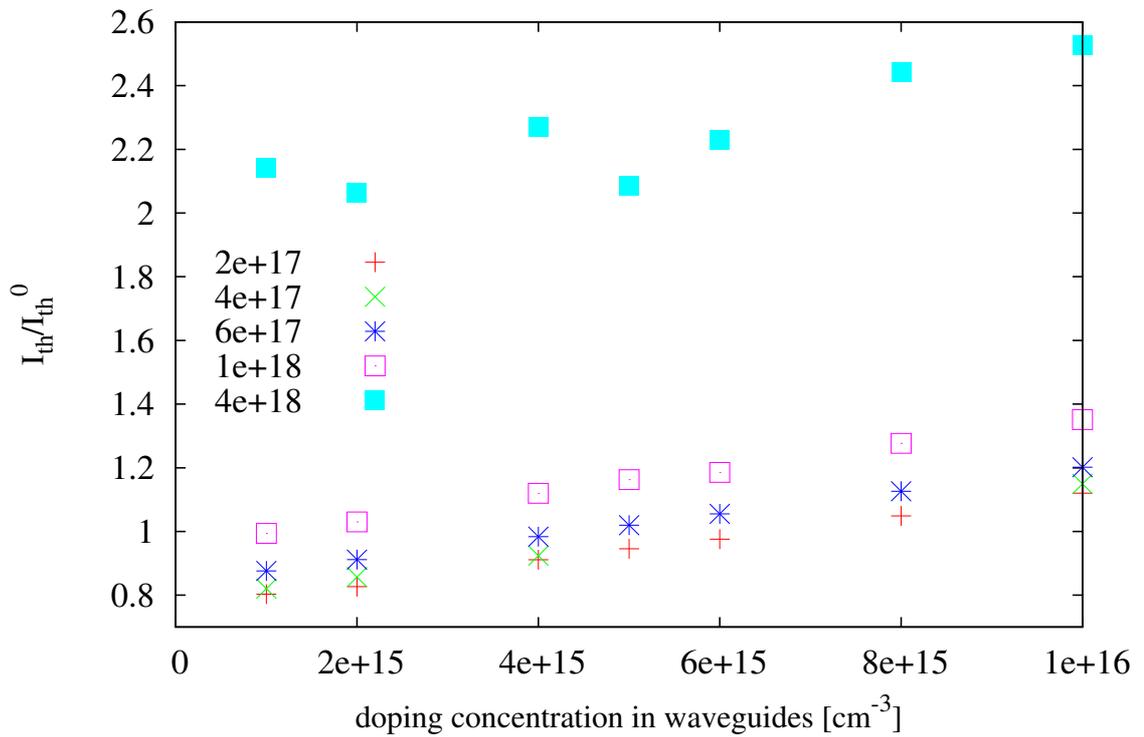}}
      \caption{Threshold current $I_{th}$, normalized by $I_{th}^0$, 
as a function of doping concentration in waveguide regions, 
when doping concentration in active region is $5\cdot 10^{14}/cm^3$, 
for several values of doping concentration in n- and p-emitters, 
as shown in the Figure. 
}
      \label{doping16c}
\end{center}
\end{figure}

\begin{figure}[t]
\begin{center}
      \resizebox{150mm}{!}{\includegraphics{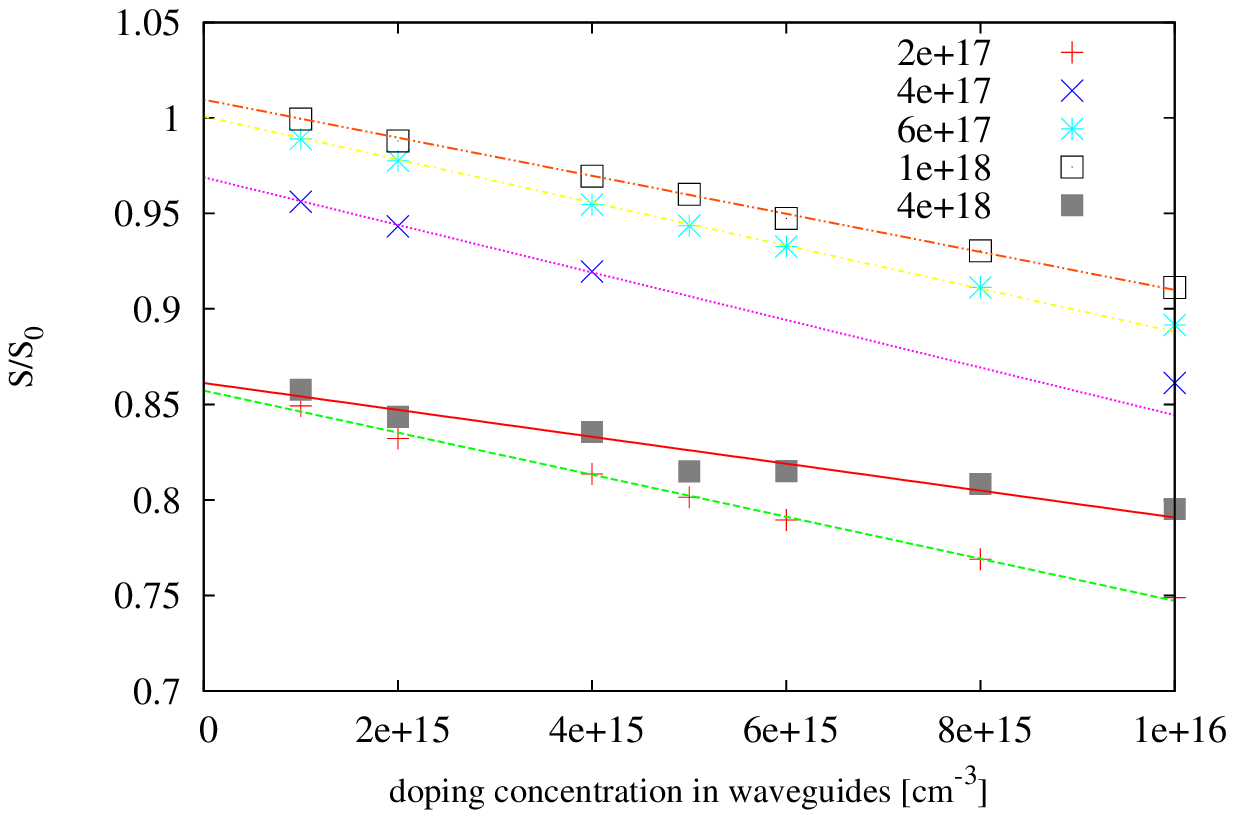}}
      \caption{The slope $S=dL/dI$, normalized by $S_0$, as a function of 
doping concentration in waveguide regions, when doping concentration in active region is 
$5\cdot 10^{14}/cm^3$, for several values of doping concentration in n- and p-emitters, 
as shown in the Figure. 
}
      \label{doping16}
\end{center}
\end{figure}

\begin{figure}[t]
\begin{center}
      \resizebox{150mm}{!}{\includegraphics{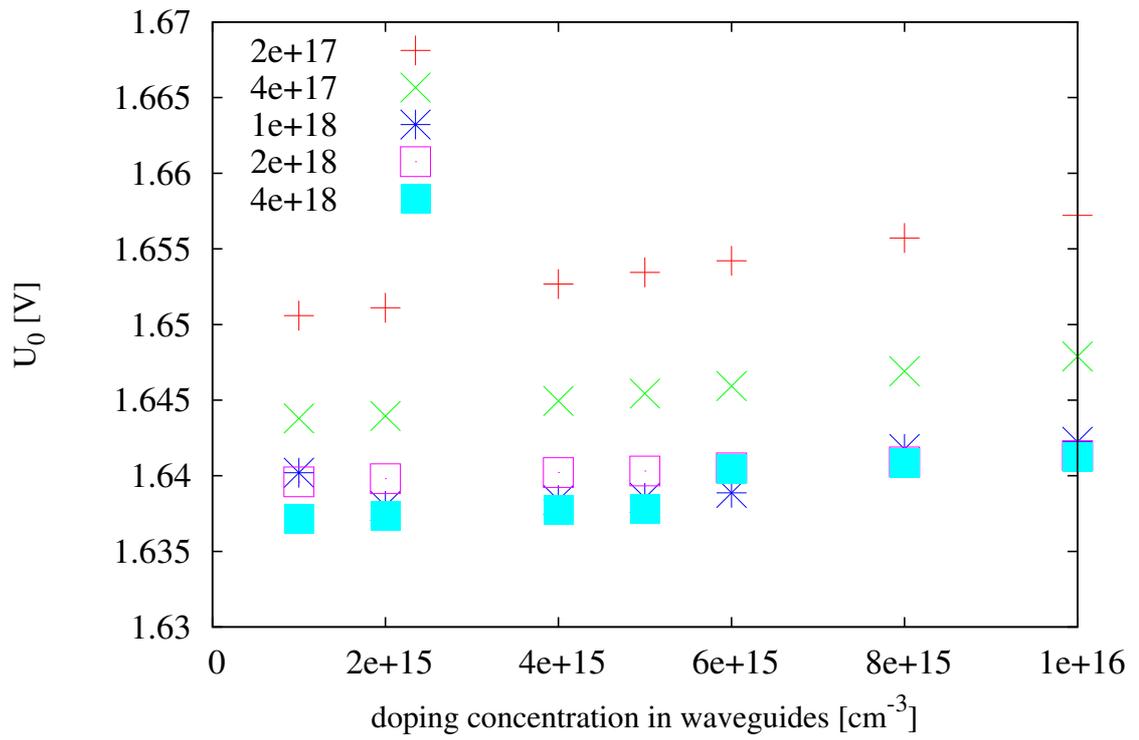}}
      \caption{Lasing offset voltage $U_0$, as a function of 
doping concentration in waveguide regions, when doping concentration in active region is 
$5\cdot 10^{14}/cm^3$, for several values of doping concentrations in n- and p-emitters, 
as shown in the Figure. 
}
      \label{doping_U02}
\end{center}
\end{figure}

\begin{figure}[t]
\begin{center}
      \resizebox{150mm}{!}{\includegraphics{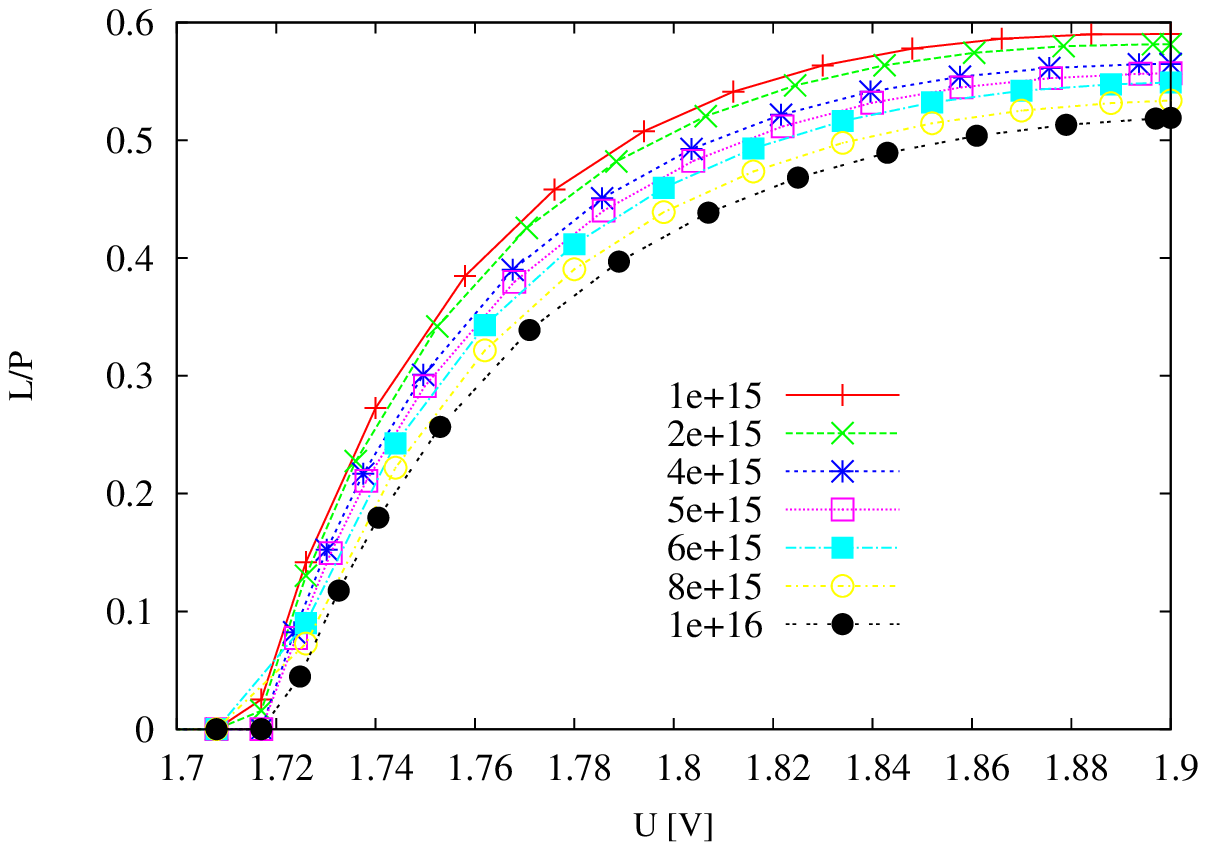}}
      \caption{Optical efficiency versus applied voltage, for a few values
		of doping concentration in waveguide region, as described in the Figure.
		Doping concentrations in active region is $10^{15}/cm^3$, and in emitters
		regions it is $10^{18}/cm^3$.
}
      \label{kpd02}
\end{center}
\end{figure}

Figures \ref{doping16c}, \ref{doping16}, \ref{doping_U02}, and \ref{kpd02} illustrate, respectively, dependencies
of threshold currend, $dL/dI$, lasing offset voltage $U_0$, and optical efficiency on doping levels in waveguide
regions. We observe, again, that the best characteristics (lowest $I_{th}$ and highest $dL/dI$ values) are at
lowest doping levels in waveguides. $U_0$ also weekly depends on doping in waveguides. The most significant
role is played by doping in emitters regions, rather. However, as Figure \ref{kpd02} shows, optical efficiency 
changes significantly with doping levels. One would like here to use Equation \ref{kpd_equation_u} to describe
the data in this Figure. However, it could provide only a very qualitative fiting of the data. Since all the data
in this figure indicate on the same or very close values of $U_0$ (compare also results of Figure \ref{doping_U02}),
we attribute large differences between these curves to significant increase of differential resistance $r=dU/dI$
when doping level increases. And that could be understood as building up of p-n junction outside of quantum well region,
with increase of doping.

\begin{figure}[t]
\begin{center}
      \resizebox{150mm}{!}{\includegraphics{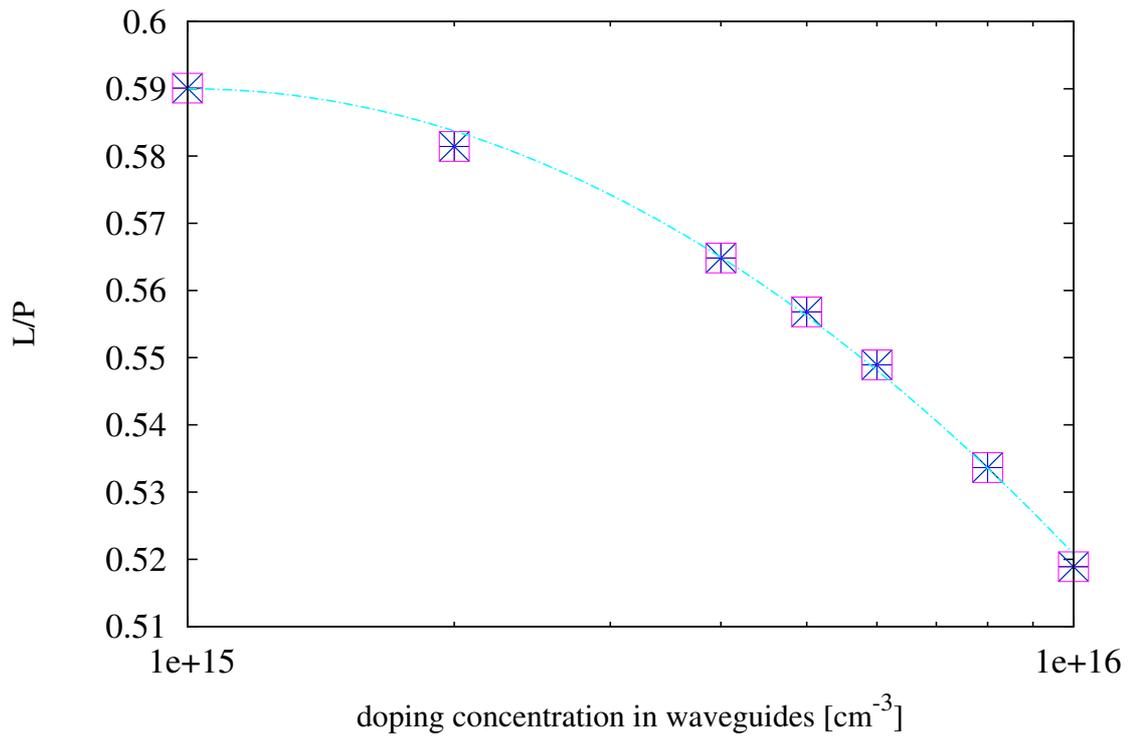}}
      \caption{Optical efficiency at applied voltage of $1.9 V$, 
		as a function of doping concentration in waveguide regions,
		when doping concentrations in active regions is $10^{15}/cm^3$, and in 
		emitters regions it is $10^{18}/cm^3$. Notice logarithmic scale on x-axes.
		The solid curve is a least-squares fiting of the function 
		$L/P = A \cdot (log(x/x_0))^2 + B$, where $A= 0.01305$, $x_0=1.0\cdot 10^{15}$,
		and $B=0.590$.
}
      \label{kpd04}
\end{center}
\end{figure}

A characteristic for "N-N" structure is parabolic dependence of optical efficiency on doping levels, as shown in Figure \ref{kpd04}.

\clearpage

\subsection{The role of doping in Emitters Regions.}

Doping levels in emitters regions contribute in the most prominent degree to laser characteristics.

\begin{figure}[t]
\begin{center}
      \resizebox{150mm}{!}{\includegraphics{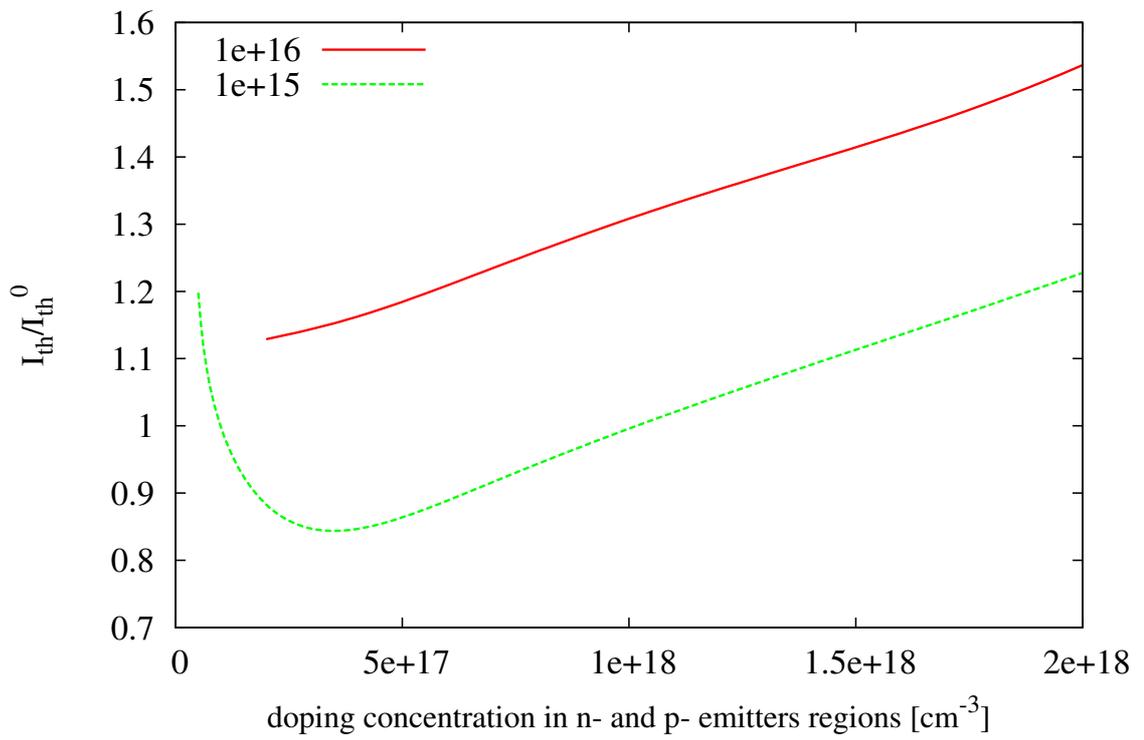}}
      \caption{Threshold current, $I_{th}$, normalized by $I_{th}^0$, as a function of 
doping concentration in n- and p-emitters, when doping concentration in active region is 
$10^{15}/cm^3$ and in waveguide $10^{15}/cm^3$ and $10^{16}/cm^3$. 
}
      \label{doping11}
\end{center}
\end{figure}

\begin{figure}[t]
\begin{center}
      \resizebox{150mm}{!}{\includegraphics{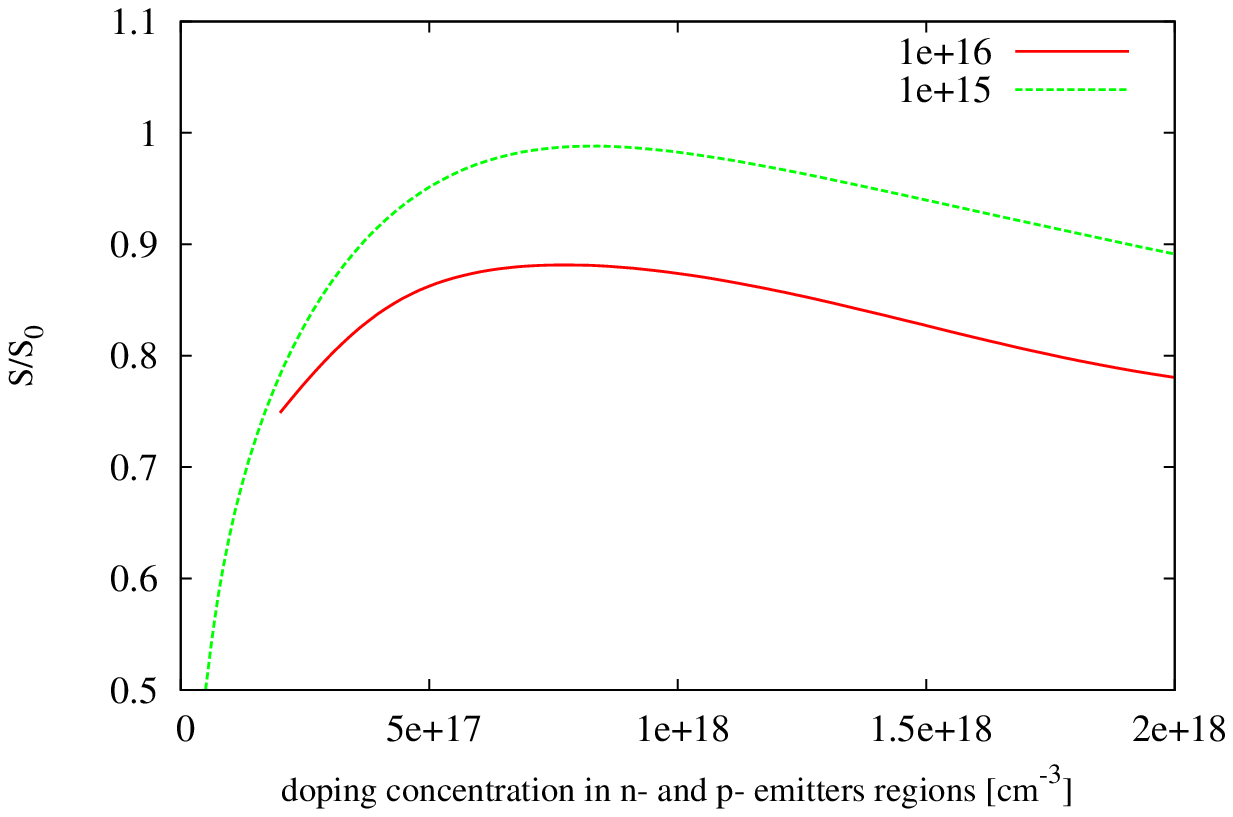}}
      \caption{The slope $S=dL/dI$, normalized by $S_0$, as a function of 
doping concentration in n- and p-emitters, when doping concentration in active region is 
$10^{15}/cm^3$ and in waveguide $10^{15}/cm^3$ and $10^{16}/cm^3$. 
}
      \label{doping12}
\end{center}
\end{figure}

Figures \ref{doping11} and \ref{doping12} show typical example dependencies of $I_{th}$ and $dL/dI$ on doping
in emitters. Clearly is observed a minimum in $I_{th}(d_e)$ at $d_e$ somewhat lower than $10^{18} cm^{-3}$,
and a maximum in $S(d_e)$ at a value of $d_e$ close to that but not the same.

\begin{figure}[t]
\begin{center}
      \resizebox{150mm}{!}{\includegraphics{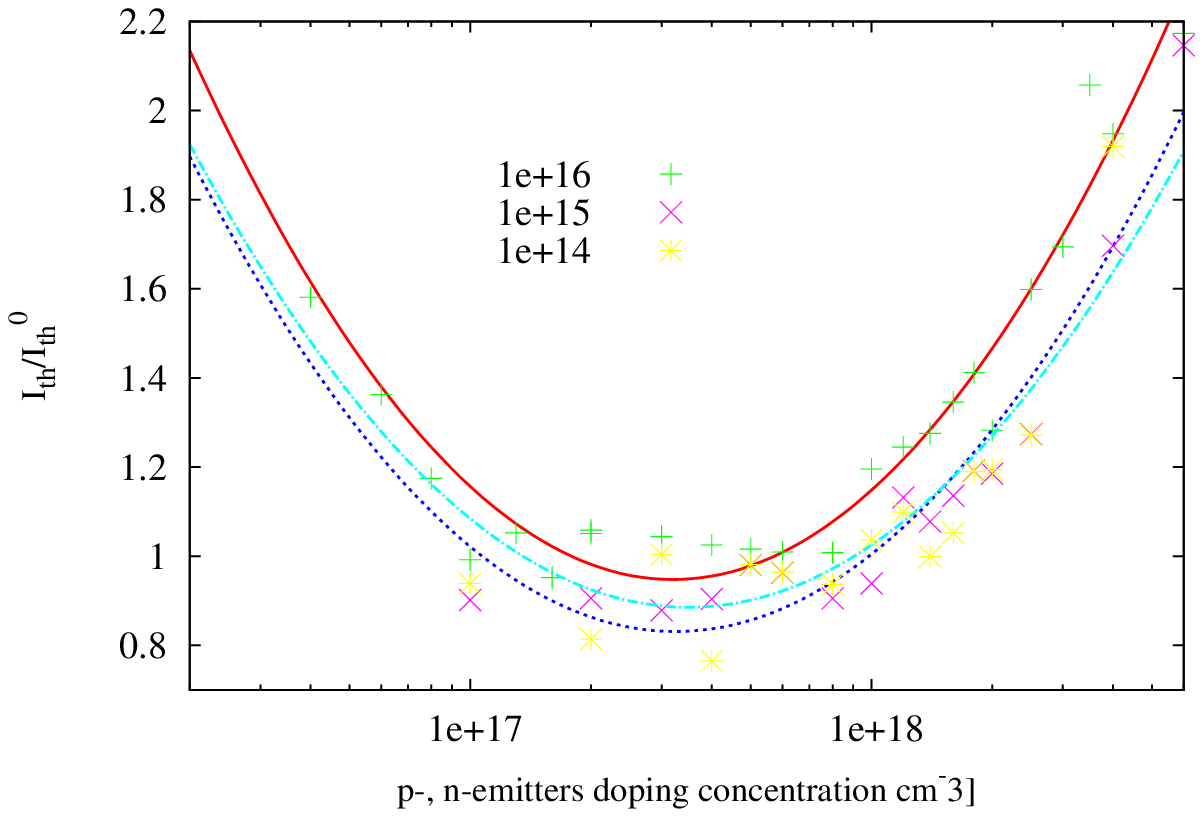}}
      \caption{Normalized threshold current as a function 
		of n- and p-emitters doping concentration, for a few values 
		of doping concentrations in active region, as indicated in the Figure.
		Doping concentration in waveguide is $10^{15}/cm^3$.
		The lines are drawn by using least-squares fiting of the data according to equation:
		$I_{th}(x) = I_0+ a \cdot (log(x/x_0))^2$, where $I_0$, $a$, and $x_0$ are certain 
		fiting parameters. In this case, for active region doping concentration of
		$10^{16}/cm^{-3}$, the parameters values are: $I_0=0.948$, $a=0.154$, $x_0= 3.2e+17$.
		When active region is $10^{15}/cm^{-3}$, 
		the parameters values are: $I_0=0.831$, $a=0.137$, $x_0= 3.25e+17$.
		When active region is $10^{14}/cm^{-3}$, 
		the parameters values are: $I_0=0.885$, $a=0.127$, $x_0= 3.5e+17$.
}
      \label{doping41}
\end{center}
\end{figure}

\begin{figure}[t]
\begin{center}
      \resizebox{150mm}{!}{\includegraphics{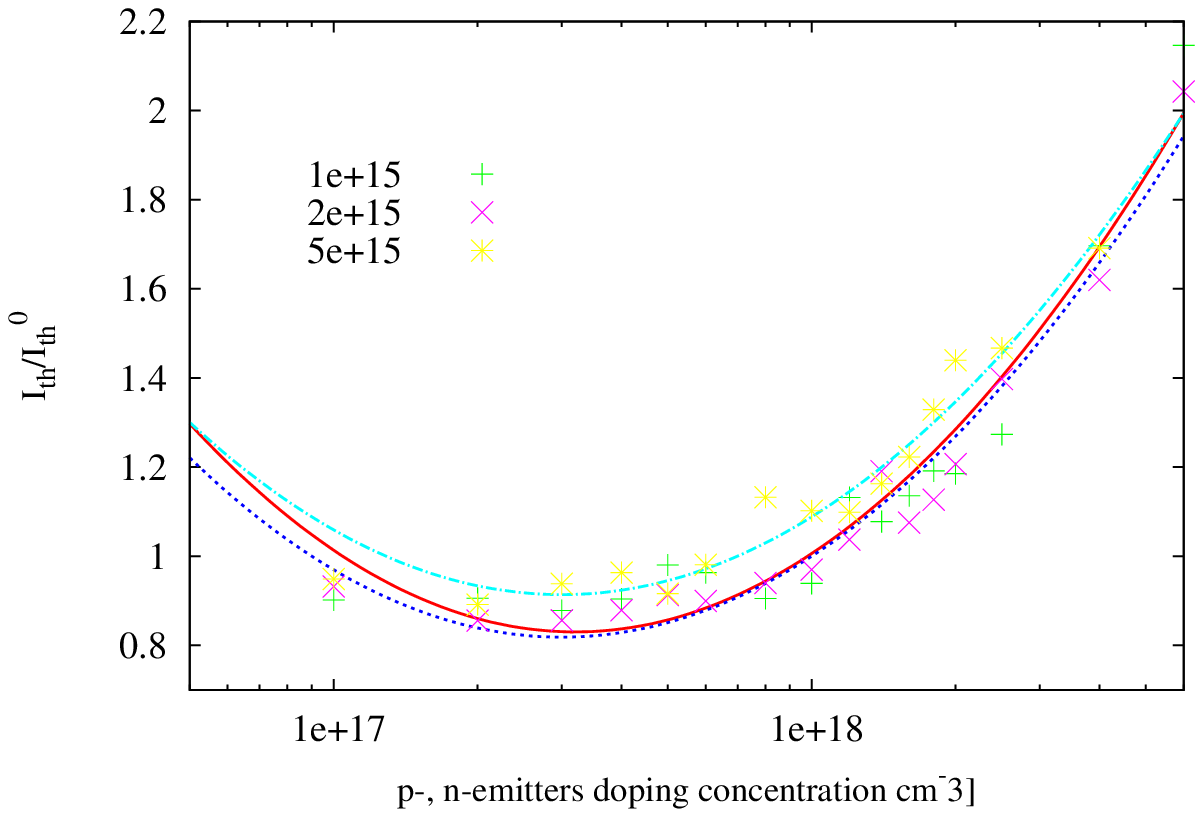}}
      \caption{Normalized threshold current as a function 
		of n- and p-emitters doping concentration, for a few values 
		of doping concentrations in waveguide regions, as indicated in the Figure.
		Doping concentration in active region is $10^{15}/cm^3$.
		The lines are drawn by using least-squares fiting of the data according to equation:
		$I_{th}(x) = I_0+ a \cdot (log(x/x_0))^2$, where $I_0$, $a$, and $x_0$ are certain 
		fiting parameters. 
}
      \label{doping43}
\end{center}
\end{figure}

\begin{figure}[t]
\begin{center}
      \resizebox{150mm}{!}{\includegraphics{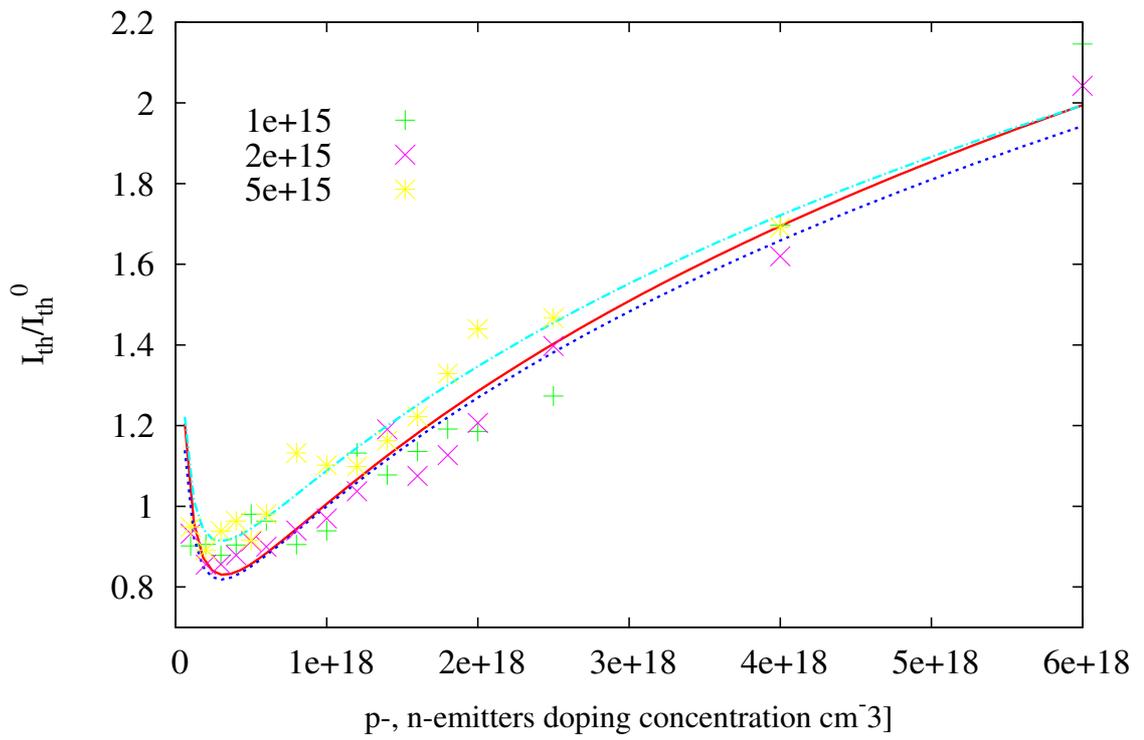}}
      \caption{The same data and lines as in Figure \ref{doping43}, except the 
		n- and p-emitters doping concentration axis is linear this time.
}
      \label{doping44}
\end{center}
\end{figure}

In Figures \ref{doping41}, \ref{doping43} and \ref{doping44}, we show the presence of such a minimum in 
$I_{th}(d_e)$ for a few more sets of data. We show that a parabolic dependence (when doping concentration is
drawn on logarythmic axes) is a good approximation of the data: compare the same results as in Figure \ref{doping43}
shown however in linear scale in Figure \ref{doping44}: 

\begin{equation}\label{parabolic_ith}
\begin{array}{ll}
		I_{th}(d_e) = I_0+ a \cdot log\left( \frac{d_e}{x_0} \right)^2, 
\end{array}
\end{equation}

where $I_0$, $a$, and $x_0$ are certain fiting parameters. 

\begin{figure}[t]
\begin{center}
      \resizebox{150mm}{!}{\includegraphics{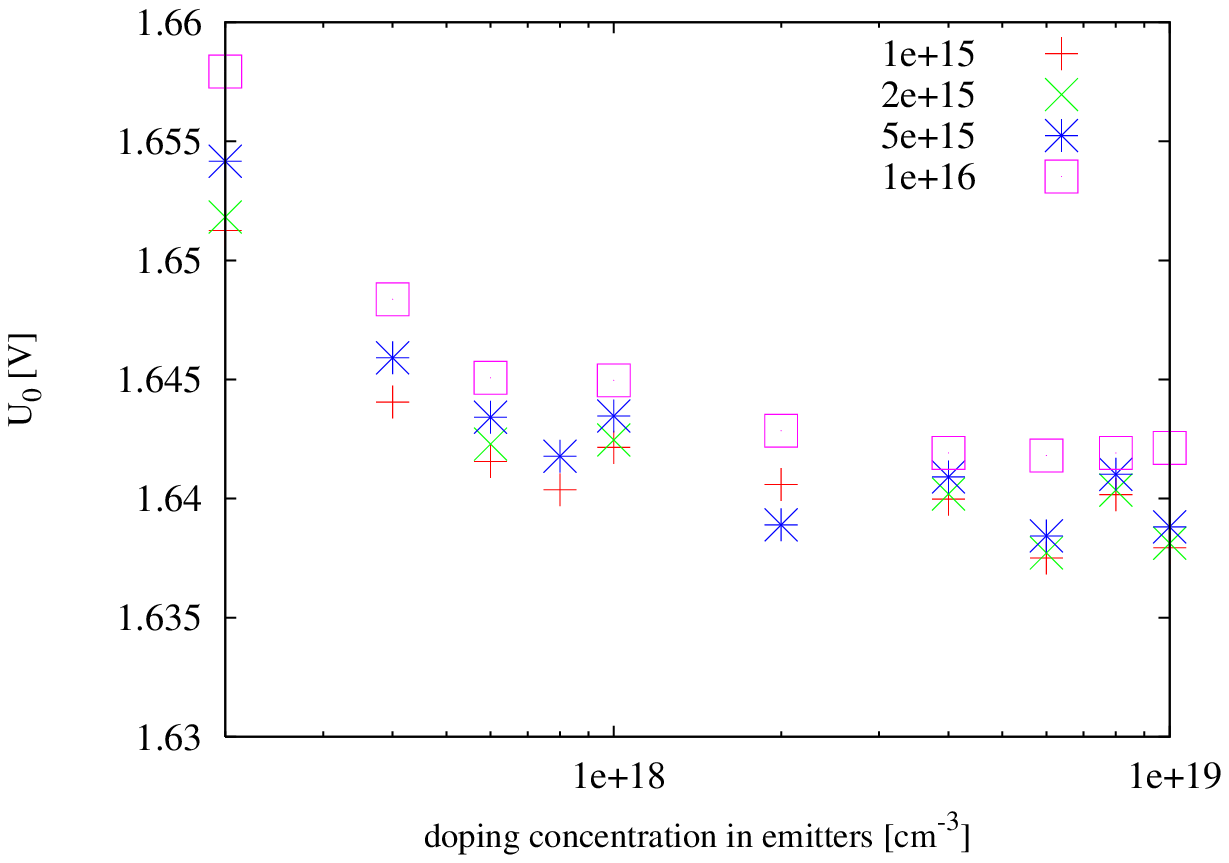}}
      \caption{$U_0$ as a function 
		of n- and p-emitters doping concentration, for a few values 
		of doping concentrations in waveguide regions, as indicated in the Figure.
		Doping concentration in active region is $10^{15}/cm^3$.
}
      \label{doping_U04}
\end{center}
\end{figure}

\begin{figure}[t]
\begin{center}
      \resizebox{150mm}{!}{\includegraphics{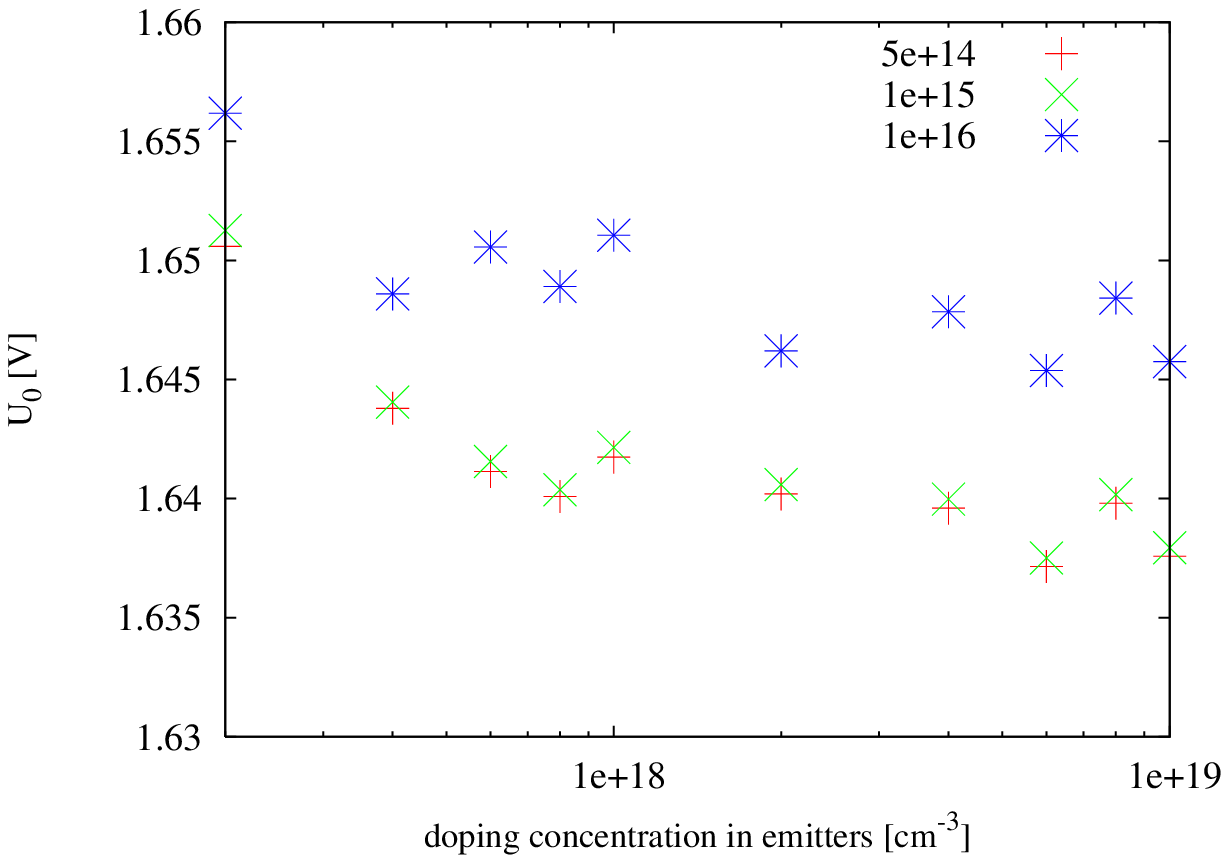}}
      \caption{$U_0$ as a function 
		of n- and p-emitters doping concentration, for a few values 
		of doping concentrations in active region, as indicated in the Figure.
		Doping concentration in wavequide regions is $10^{15}/cm^3$.
}
      \label{doping_U03}
\end{center}
\end{figure}

The lasing offset voltage $U_0$ (Figures \ref{doping_U04} and \ref{doping_U03}) 
is sensitive to emitters doping at low levels, only, below around $10^{18} cm^{-3}$.

\begin{figure}[t]
\begin{center}
      \resizebox{150mm}{!}{\includegraphics{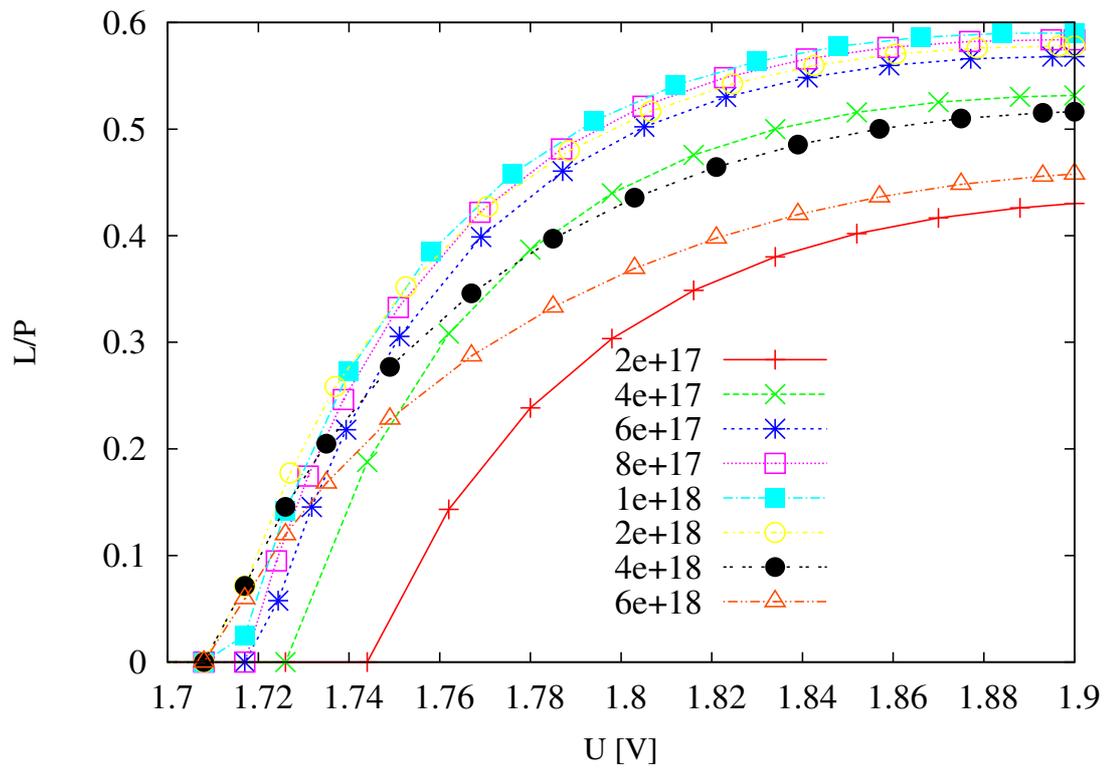}}
      \caption{Optical efficiency versus applied voltage, for a few values
		of n- and p-emitters doping concentration, as described in the Figure.
		Doping concentrations in active region and waveguide regions is $10^{15}/cm^3$.
}
      \label{kpd00}
\end{center}
\end{figure}

Strong dependencies of $I_{th}$ and $dL/dI$ on doping manifest itself in a strong dependence
of optical efficiency on doping levels, as Figure \ref{kpd00} shows. Again, it has 
a parabolic character (Figure \ref{kpd00}), similar to that one found for waveguides (Figure \ref{kpd04}).

\begin{figure}[t]
\begin{center}
      \resizebox{150mm}{!}{\includegraphics{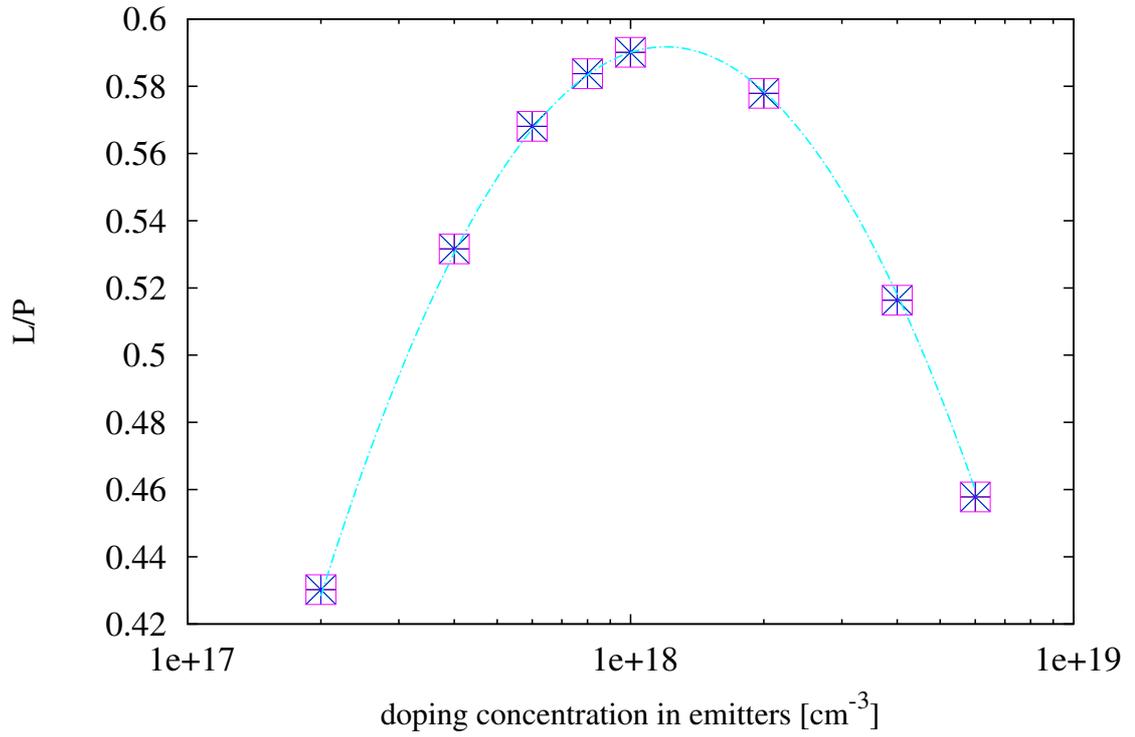}}
      \caption{Optical efficiency at applied voltage of $1.9 V$, 
		as a function of doping concentration in emitters regions,
		when doping concentrations in waveguide and active regions is $10^{15}/cm^3$.
		The solid curve is a least-squares fiting of the function 
		$L/P = A \cdot (log(x/x_0))^2 + B$, where $A= 0.0509$, $x_0=1.2\cdot 10^{18}$,
		and $B=0.5917$.
}
      \label{kpd03}
\end{center}
\end{figure}

\clearpage

\section{N-P waveguide structure.}

By "N-P" waveguide structure we mean a doping structure that differes
from that described in Table \ref{table_1},
where both waveguides and the active region (QW) are n-type doped. 
Here we assume that the waveguide on the side of n-emitter is of n-type, 
the other one is of p-type, and QW doping is of n-type.

The most remarkable difference between N-N and N-P structures is that in case of the last one 
a large reduction of $I_{th}$ and a significant increase in $dL/dI$ and in optical efficiency are
found.

\clearpage

\subsection{Active region doping}.

\begin{figure}[t]
\begin{center}
      \resizebox{150mm}{!}{\includegraphics{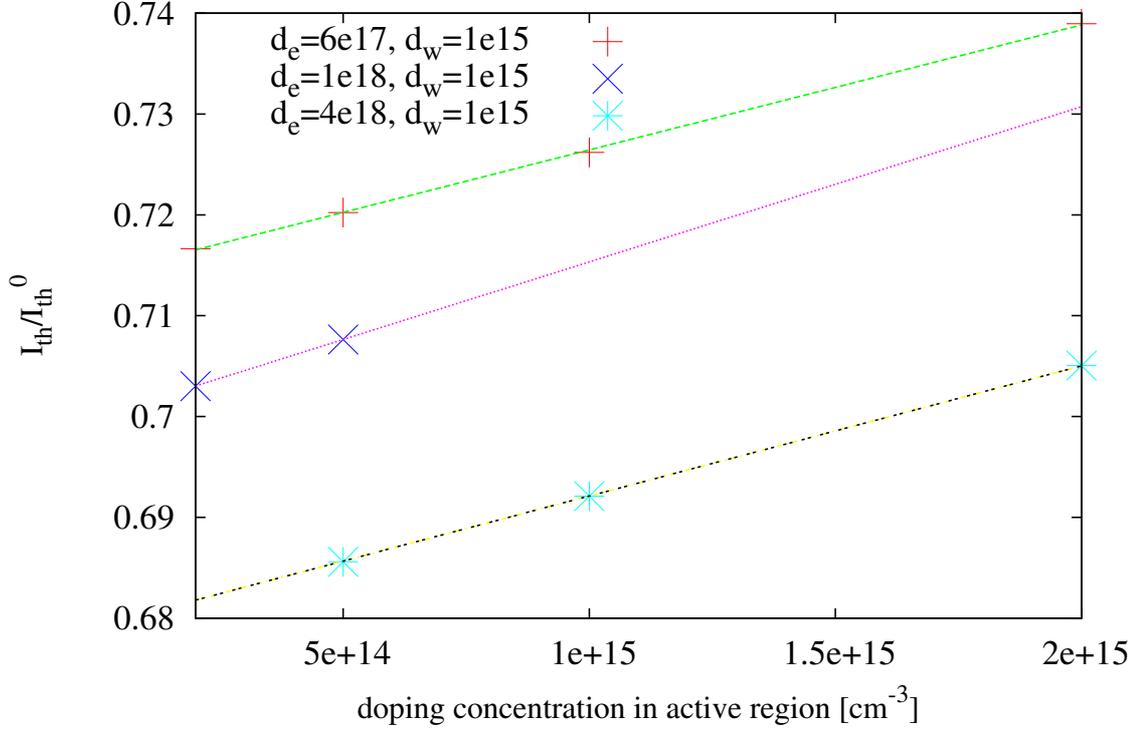}}
      \caption{Threshold current (normalized by $I_{th}^0$)
		as a function of carrier doping concentrations in active region, for a few values of concentration 
		in emitters. Concentrations in waveguide regions is $10^{15} / cm^3$.
}
      \label{np06}
\end{center}
\end{figure}

Figures \ref{np06} and \ref{np06} show the role of 
doping concentration in active region, for a few values of doping concentrations 
in emitter regions, on basic characteristics of lasers: threshold current and slope of 
light power versus current, $dL(I)/dI$, respectively. 

We see that none of these quantities depends strongly on active region doping. These effects
are similar to the onse observed for "N-N" type of waveguide structure.


\begin{figure}[t]
\begin{center}
      \resizebox{150mm}{!}{\includegraphics{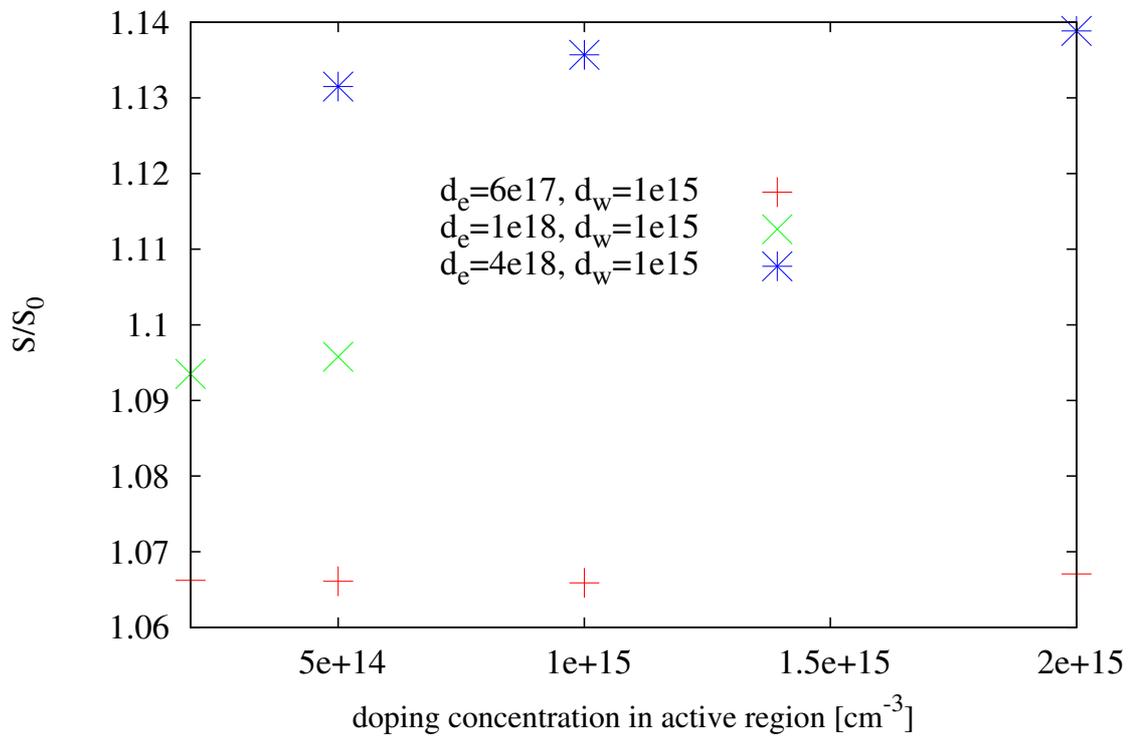}}
      \caption{Slope of $dL/dI$ (normalized by $S_0$) for the data corresponding to these in Figure \ref{np06}.
}
      \label{np07}
\end{center}
\end{figure}

\clearpage

\subsection{Waveguides doping}.

The effects of doping in waveguide regions are much stronger for N-P structure 
than these reported for N-N one.

\begin{figure}[t]
\begin{center}
      \resizebox{150mm}{!}{\includegraphics{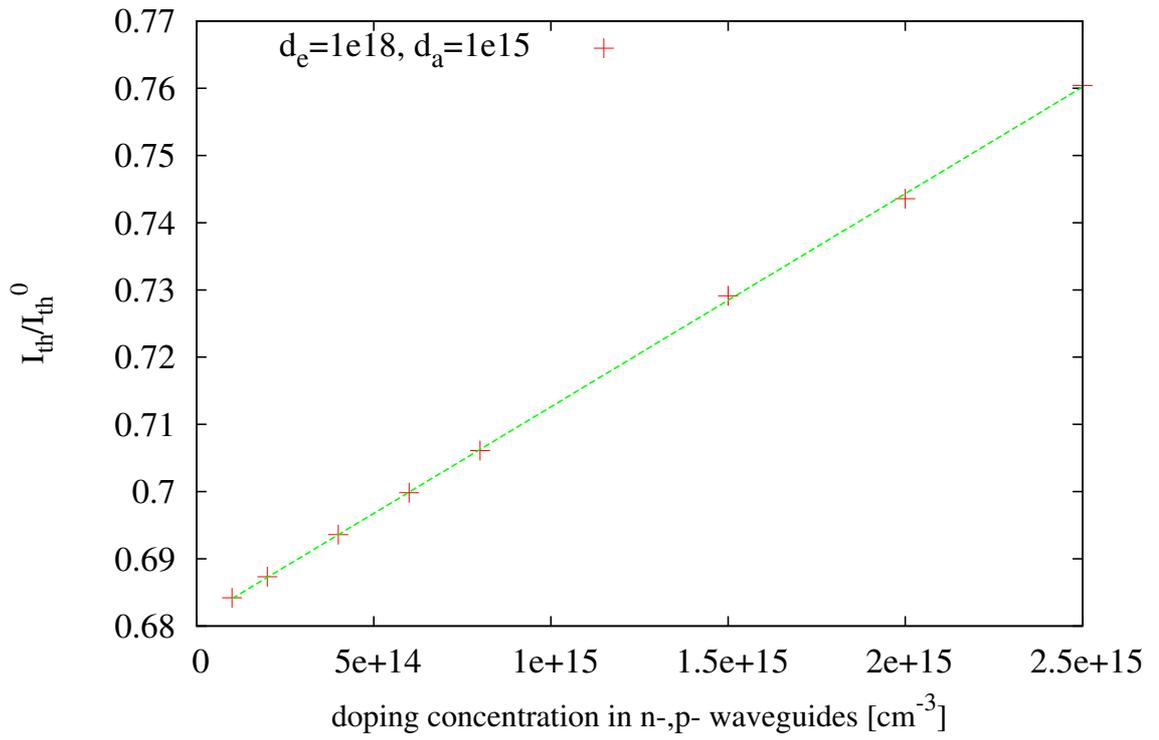}}
      \caption{Threshold current (normalized by $I_{th}^0$)
		as a function of carrier doping concentrations in waveguides.
		Concentrations in emitter regions are $10^{18} / cm^3$ (n- and p-), and in active region
		it is $n=10^{15} / cm^3$. The straigt line is given by function 
		$f(x)= 3.17 \cdot 10^{-17} \cdot x + 0.681$.
}
      \label{np02}
\end{center}
\end{figure}

Both, lasing threshold current (Figure \ref{np02}) and corresponding $dL/dI$ (Figure \ref{np02}) 
show a nice strong linear dependence on waveguides doping. Since the first one has a positive slope 
and the second one negative, there will no maximum as a function of $d_w$ in this case, which is 
different that for the case of N-N structure.

\begin{figure}[t]
\begin{center}
      \resizebox{150mm}{!}{\includegraphics{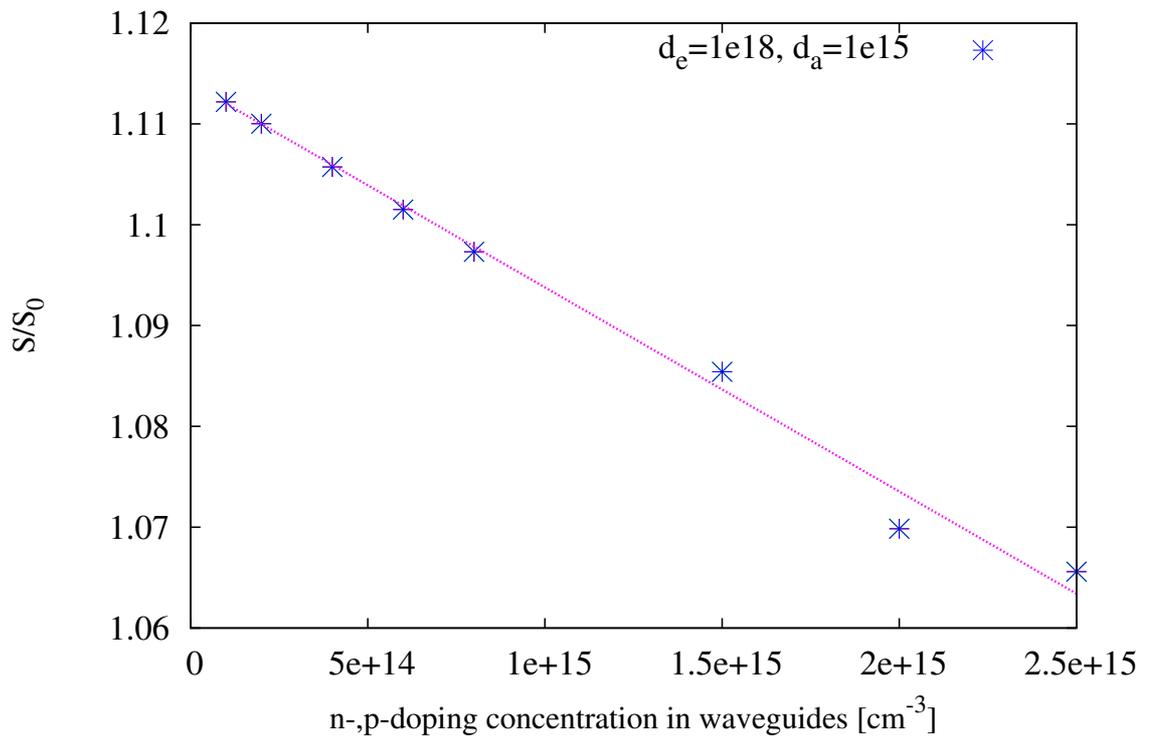}}
      \caption{The slope of Optical Power, dL/dI,
		as a function of carrier doping concentrations in waveguides.
		Concentrations in emitter regions are $10^{18} / cm^3$ (n- and p-), and in active region
		it is $n=10^{15} / cm^3$. The straigt line is given by function 
		$f(x)= -2.54 \cdot 10^{-17} \cdot x + 1.397$.
}
      \label{np03}
\end{center}
\end{figure}

The lasing offset voltage $U_0$ corresponding to the data displayed in Figures \ref{np02} and \ref{np03}
does not change with waveguide doping (within the accuracy of our modeling, and the studied doping range), 
and it is of the value of $1.711 V$.

\begin{figure}[t]
\begin{center}
      \resizebox{150mm}{!}{\includegraphics{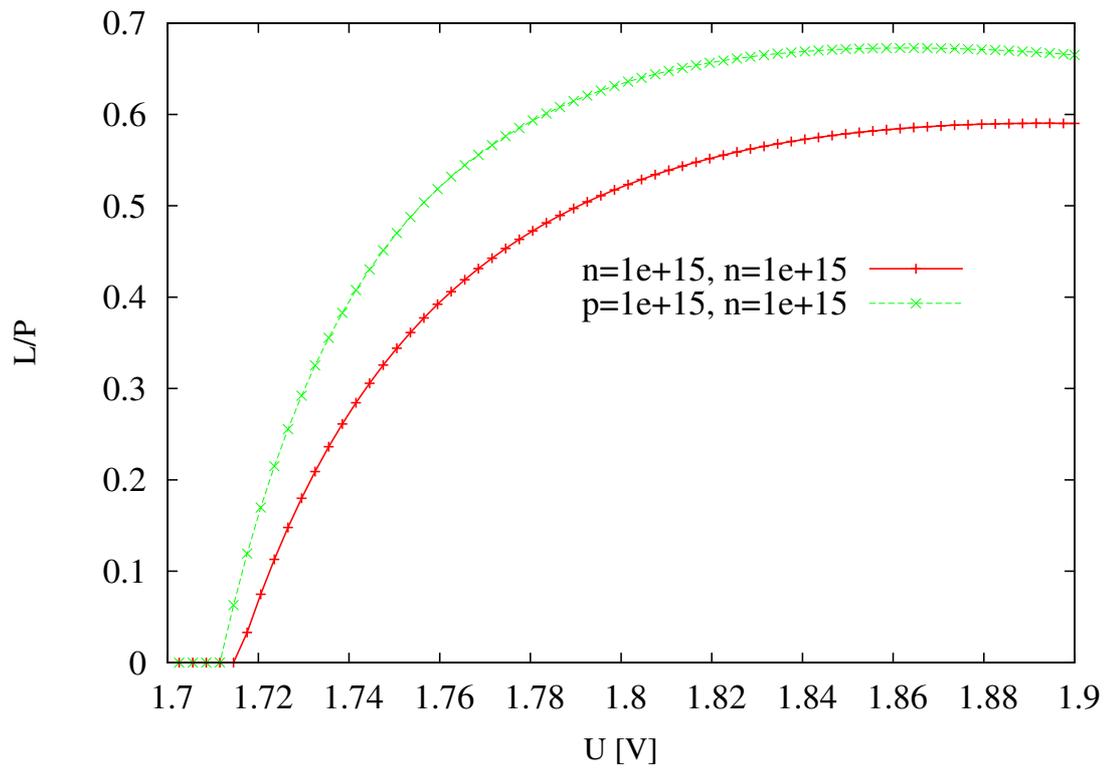}}
      \caption{Optical efficiency as a function of voltage,
		for a few combinations of carrier doping concentrations in waveguides,
		as described in the Figure.
		Concentrations in emitter regions are $10^{18} / cm^3$ (n- and p-), and in active region
		it is $n=10^{15} / cm^3$. 
}
      \label{np01a}
\end{center}
\end{figure}

In Figure \ref{np01a} we compare optical efficiency for both types of structures.

\clearpage

\subsection{Doping in Emitters}.

As in case on N-N structures, here also doping levels in emitters regions 
contribute the most to laser characteristics.

\begin{figure}[t]
\begin{center}
      \resizebox{150mm}{!}{\includegraphics{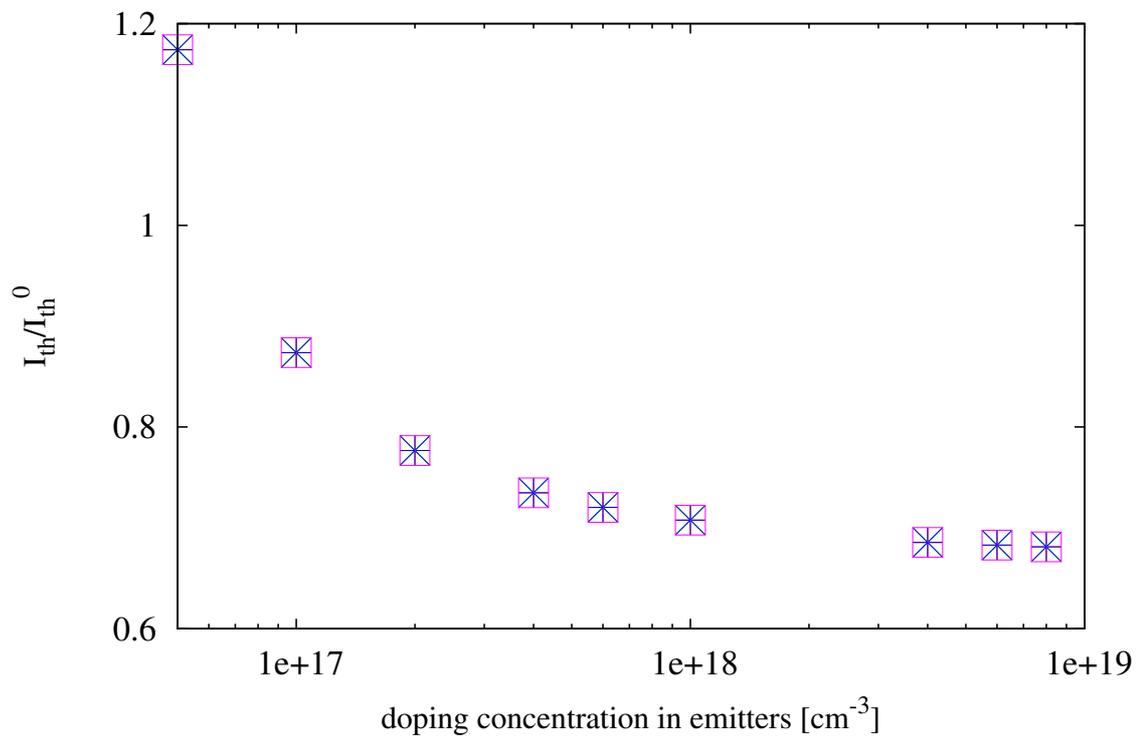}}
      \caption{Threshold current as a function of carrier doping concentrations in emitters.
		Concentrations in waveguide and active regions are $10^{15} / cm^3$
}
      \label{np31}
\end{center}
\end{figure}

Figures \ref{np31} and \ref{np31} show example dependencies of $I_{th}$ and $dL/dI$ on doping
in emitters. There are no maxima/minima there like in case
of N-N structure, and at concentrations above about $10^{18} cm^{-3}$ the effect of doping becomes small.

The lasing offset voltage $U_0$ (Figure \ref{np09}) 
is sensitive to emitters doping at low levels, only, below around $10^{18} cm^{-3}$. 
This is the same as for N-N structure.

\begin{figure}[t]
\begin{center}
      \resizebox{150mm}{!}{\includegraphics{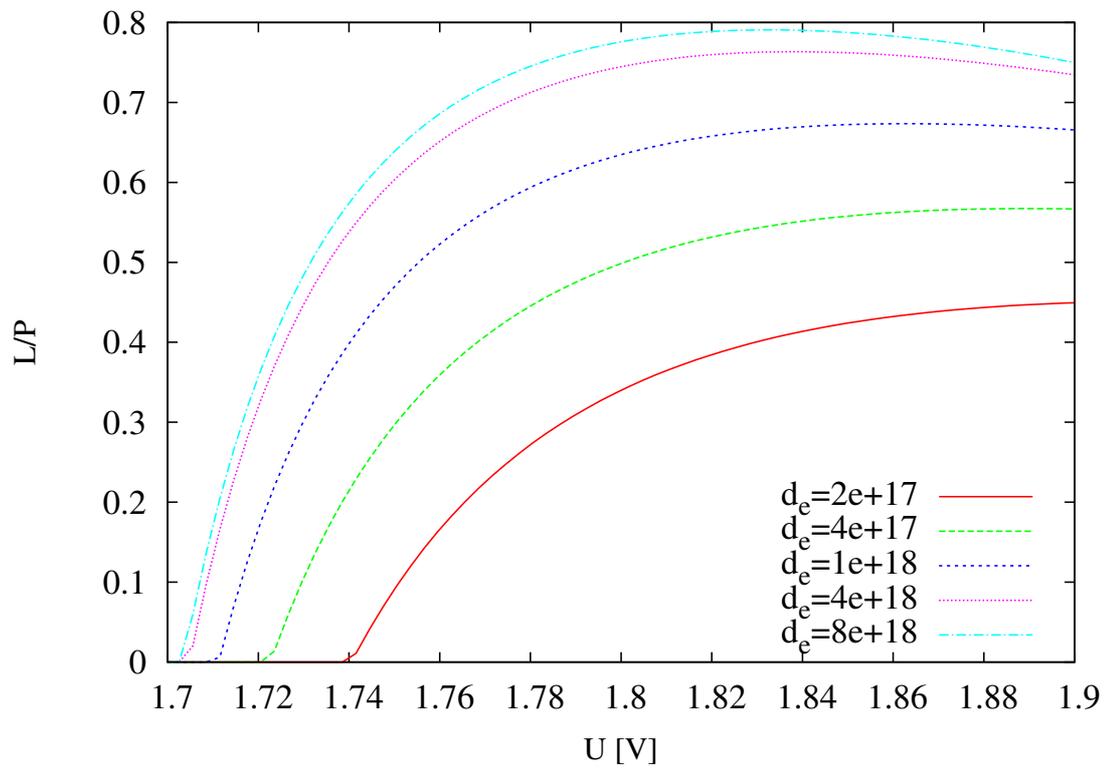}}
      \caption{Optical efficiency for a few values of doping concentration in emitters.
		Concentration in waveguide regions is $10^{15} / cm^3$ (n- and p-), and in active region
		it is $n=5\cdot 10^{14} / cm^3$. 
}
      \label{np08}
\end{center}
\end{figure}

\begin{figure}[t]
\begin{center}
      \resizebox{150mm}{!}{\includegraphics{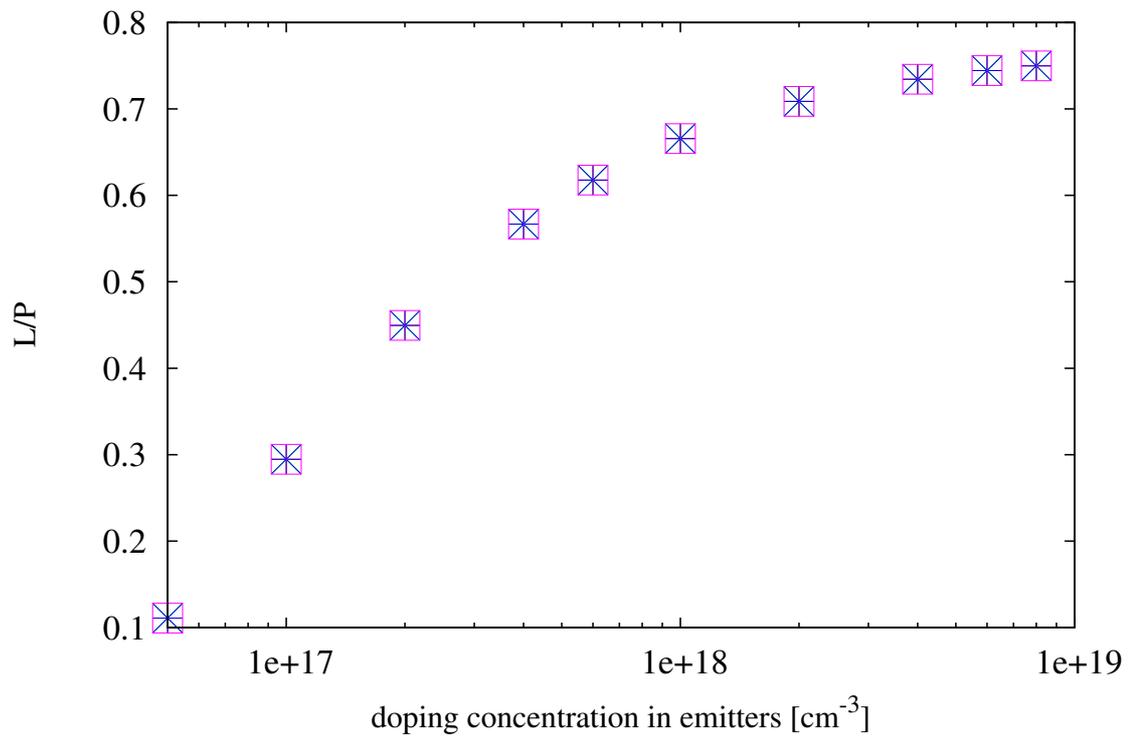}}
      \caption{Optical efficiency at applied voltage of $1.9 V$, 
		as a function of doping concentration in emitters regions,
		when doping concentrations in waveguide is $10^{15}/cm^3$ and in active regions it is $5\cdot 10^{14}/cm^3$.
}
      \label{kpd29}
\end{center}
\end{figure}

Optical efficiency does depend weekly on doping at large doping levels, unlike for N-N structures
(Figures \ref{np08} and \ref{kpd29}).

\begin{figure}[t]
\begin{center}
      \resizebox{150mm}{!}{\includegraphics{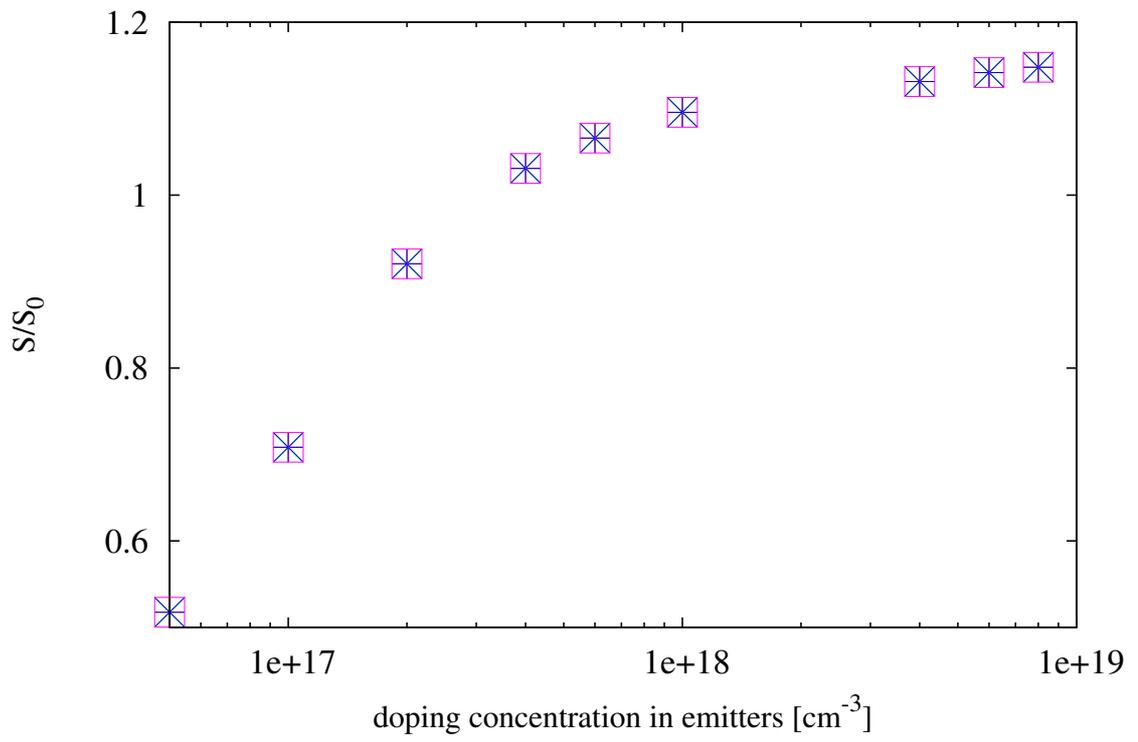}}
      \caption{Slope $dL/dI$ as a function of carrier doping concentrations in emitters.
		The data correspond to these in Figures \ref{np31} and \ref{np09}.
}
      \label{np32}
\end{center}
\end{figure}

\begin{figure}[t]
\begin{center}
      \resizebox{150mm}{!}{\includegraphics{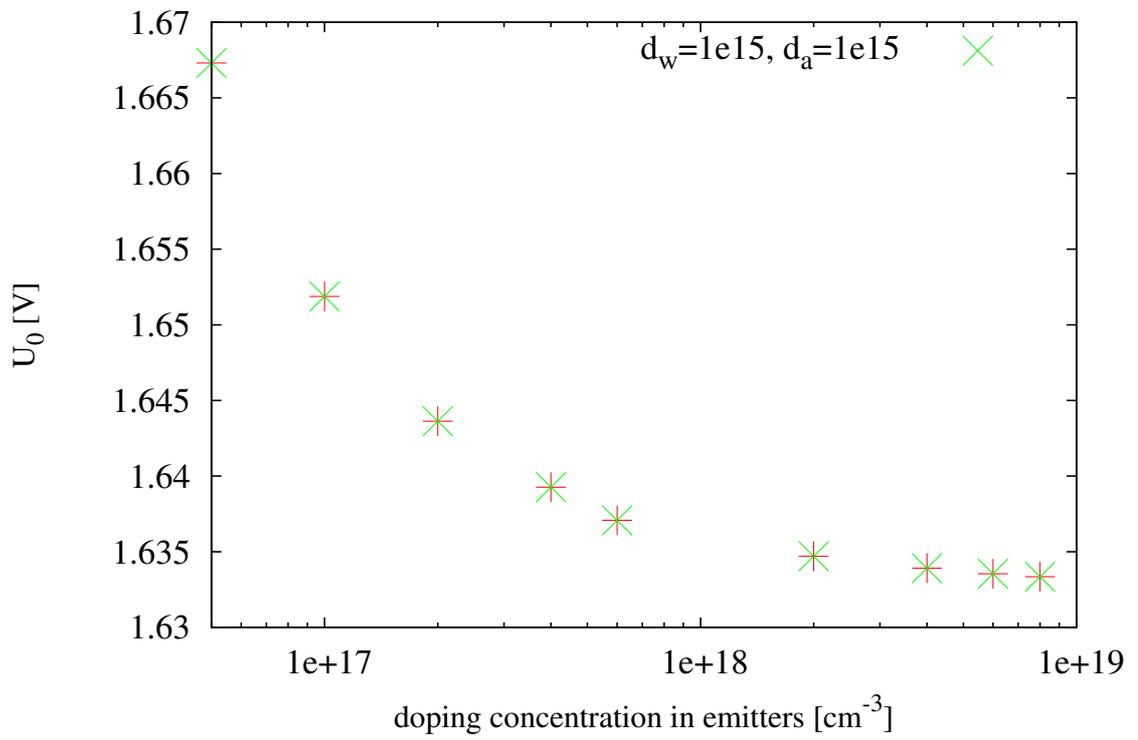}}
      \caption{Lasing offset voltage as a function of carrier doping concentrations in emitters.
		The data correspond to these in Figures \ref{np31} and \ref{np32}.
}
      \label{np09}
\end{center}
\end{figure}

\clearpage
\section{Summary and Conclusions.}

SCH AlGaAs lasers with two kinds of waveguide doping structures have been modeled with Synopsys 
TCAD: "N-N" type structure where both waveguides have n-type of doping (with QW doping of n-type
as well), and "N-P" structure where the waveguide on the side of p-emitter has p-type doping. 
The "N-N" structure is of the type we have experimental data for and it was used as a reference
in our calculation calibration. 

We observed that the doping level in active region has a small only influence on laser characteristics
(lasing threshold current, slope of light power versus current, lasing offset voltage, optical efficiency). 
Better results are obtained for the lowest possible doping. 
More pronounced influence is of doping in waveguide regions. For "N-N" type of structure 
an optimal doping level is of around $10^{15} cm^{-3}$, but for "N-P" type of structure,
 it is desirable to have lower level of doping concentration there.

The most significant is the role of doping levels in emitters. It should be around $10^{18} cm^{-3}$
in case of both types of structures. However, "N-P" type of structure
gives significantly better results than "N-N" one, with threshold current lowered to about 70\%
and optical efficiency increased to near 80\%.

\clearpage

\end{document}